\documentclass[superscriptaddress,aps,pra,onecolumn,showkeys]{revtex4-2}
\usepackage[colorlinks,linkcolor=blue,anchorcolor=blue,citecolor=blue,urlcolor=blue]{hyperref}
\usepackage{amsmath,mathtools,yhmath,siunitx}
\usepackage{amssymb}
\usepackage{graphicx}
\usepackage{color}
\usepackage{bm}

\begin{document}
\title{Highly Scalable Quantum Router with Frequency-Independent Scattering Spectra}

\author{Yue Cai}
\affiliation{Lanzhou Center for Theoretical Physics, Key Laboratory of Theoretical Physics of Gansu Province, and Key Laboratory of Quantum Theory and Applications of MoE, Lanzhou University, Lanzhou, Gansu 730000, China}

\author{Kang-Jie Ma}
\affiliation{Lanzhou Center for Theoretical Physics, Key Laboratory of Theoretical Physics of Gansu Province, and Key Laboratory of Quantum Theory and Applications of MoE, Lanzhou University, Lanzhou, Gansu 730000, China}

\author{Jie Liu}
\affiliation{Lanzhou Center for Theoretical Physics, Key Laboratory of Theoretical Physics of Gansu Province, and Key Laboratory of Quantum Theory and Applications of MoE, Lanzhou University, Lanzhou, Gansu 730000, China}

\author{ Gang-Feng Guo}
\affiliation{Lanzhou Center for Theoretical Physics, Key Laboratory of Theoretical Physics of Gansu Province, and Key Laboratory of Quantum Theory and Applications of MoE, Lanzhou University, Lanzhou, Gansu 730000, China}

\author{Lei Tan}
\email{tanlei@lzu.edu.cn}
\affiliation{Lanzhou Center for Theoretical Physics, Key Laboratory of Theoretical Physics of Gansu Province, and Key Laboratory of Quantum Theory and Applications of MoE, Lanzhou University, Lanzhou, Gansu 730000, China}

\author{Wu-Ming Liu}
\affiliation{Beijing National Laboratory for Condensed Matter Physics, Institute of Physics, Chinese Academy of Sciences, Beijing 100190, China}

\begin{abstract}
Optical quantum routers play a crucial role in quantum networks and have been extensively studied in both theory and experiment, leading to significant advancements in their performance. However, these routers impose stringent requirements for achieving desired routing results, as the incident photon frequency must be in strict resonance with one or several specific frequencies. To address this challenge, we propose an efficient quantum router scheme composed of semi-infinite coupled-resonator waveguide (CRW) and a giant atom. The single-channel router scheme enables stable output with 100\% transfer rate over the entire energy band of the CRW. Leveraging this intriguing result, we further propose a multi-channel router scheme that possesses high stability and universality, while also being capable of performing various functionalities. The complete physical explanation of the underlying mechanism for this intriguing result is also presented. We hope that quantum router with output results unaffected by the frequency of the incoming information carriers presents a more reliable solution for the implementation of quantum networks.

\end{abstract}

\maketitle

\section{INTRODUCTION}
Quantum networks represent an emerging paradigm in information processing\cite{quantum_network_1,quantum_network_2,quantum_network_3,quantum_network_4,quantum_network_5}, playing an indispensable role in the implementation of future technologies such as quantum communication\cite{quantum_communication_1,quantum_communication_2,quantum_communication_3,quantum_communication_4}, distributed quantum computing\cite{quantum_computing_1}, and quantum metrology\cite{quantum_metrology_1,quantum_metrology_2}. The optical quantum router serves as a fundamental building block in quantum networks. Its primary functionality lies in governing photons to enable on-demand routing of signals from their source to various quantum channels.\cite{information_carriers_router_1,information_carriers_router_2,1_d_crw_router,optical_router}. As a result, quantum routers, based on various structures such as linear waveguide-emitters\cite{quantum_zeno_switch,two_waveguide_atom_2,two_waveguide_atom_3}, coupled-resonator waveguides (CRWs)-atoms\cite{x_shape_router_1,x_shape_router_2,t_shape_router_1}, atomic ensembles\cite{atomic_ensemble_router}, superconducting circuits\cite{superconducting_circuit_router_1,superconducting_circuit_router_2,superconducting_circuit_router_3}, and optomechanical systems\cite{optomechanical_system_router_1,optomechanical_system_router_2,optomechanical_system_router_3}, have been extensively researched and have made significant strides. They have evolved from single-output channel devices to supporting multiple output channels\cite{optical_router,x_shape_router_1,two_semi_infinite_crw_router}, and have increased their maximum transfer rates from 50\% to 100\%\cite{t_shape_router_1,multi_coupled_points_crw_router}.
However, the aforementioned routers achieve desired routing results, including 100\% transfer rate, only when photon with some specific frequencies are input. Even a slight detuning in incident photon frequency from these specific values substantially affect the routing results. Consequently, this limitation significantly restricts the choice of photons as information carriers. Moreover, the presence of dissipation in non-ideal devices results in unavoidable energy losses during the routing process\cite{non_ideal_devices_1,non_ideal_devices_2,non_ideal_devices_3,non_ideal_devices_4}. Due to the high sensitivity of the scattering spectra of these routers to frequency, such losses will render the outputs of the routers uncontrollable, making quantum network development significantly challenging.

To address this issue, this paper presents a stable single-photon quantum routing scheme. This scheme comprises two or more semi-infinite CRWs\cite{semi_infinite_crw,multiple_semi_infinite_crw_router} and a giant atom and can be realized by superconducting quantum circuits\cite{SQC_1,SQC_2,SQC_3,SQC_4,SQC_CRW_1,SQC_CRW_2,SQC_CRW_3,SQC_GA_1,SQC_GA_2,SQC_GA_3}. The giant atom, featuring a cyclic three-level structure, acts as a quantum node in this scheme. When it interacts with the bosonic field, its size is of the same order of magnitude as the bosonic field's wavelength, preventing it from being approximated as a point-like entity. This renders the conventional electric-dipole approximation in quantum optics no longer applicable to giant atoms\cite{giant_atom_1,giant_atom_2,giant_atom_engineering,giant_atom_3,giant_atom_crw_sistem}. This unique property enables a giant atom to interact simultaneously with multiple connecting points and the interference effect between these points will dramatically modulate the collective behavior of the giant atom\cite{SQC_CRW_2,giant_atom_engineering}. Leveraging this uniqueness, the giant atom effectively replaces small atoms for modulating the on-demand output of photons into different channels. 
As a result, the giant atom has been widely employed in various router schemes in recent years, introducing novel phenomena arising from nonlocal coupling, such as non-reciprocal scattering and interference effects\cite{giant_atom_router_1,giant_atom_router_2,giant_atom_router_3}.
On the other hand, we employ semi-infinite CRW instead of infinite CRW in traditional scheme as the quantum channel. Leveraging the hard-wall boundary condition imposed on the radiation field by the termination of such semi-infinite CRW, the photon transfer rate can be significantly improved\cite{semi_infinite_crw,t_shape_router_1}. With the combined effect of the nonlocal coupling of giant atom and the boundary conditions formed by the termination of the semi-infinite CRW, our model can maintain 100\% transfer rates over the entire energy band of CRW, and demonstrate a scattering spectra independent of photon frequency. This intriguing result releases the router from constraints related to the selection of incident photon frequencies, ensuring relatively stable output even in the presence of energy losses. Such characteristics provide a more reliable solution for the further development of quantum networks.

This paper is organized as follow: In Sec. \ref{II}, we introduce our single-channel quantum router, consisting of a cyclic three-level giant atom coupled to two semi-infinite CRWs and provide the model's Hamiltonian. In Sec. \ref{III}, we employ a discrete coordinate scattering approach to investigate single-photon scattering processes, discuss the routing properties of this model, and explain the physical mechanism behind the unusual phenomenon of 100\% transfer rates over the entire energy band. In Sec. \ref{IV}, the single-channel router is expanded into a multi-channel router to achieve more versatile functionality. Additionally, we present a feasible experimental implementation of these routers based on the superconducting quantum circuit platform. Finally, a brief summary and discussion of potential future extensions are provided in the concluding section.

\section{MODEL}\label{II}

\begin{figure*}[htb]
	
	\centering
	\includegraphics[width=15cm]{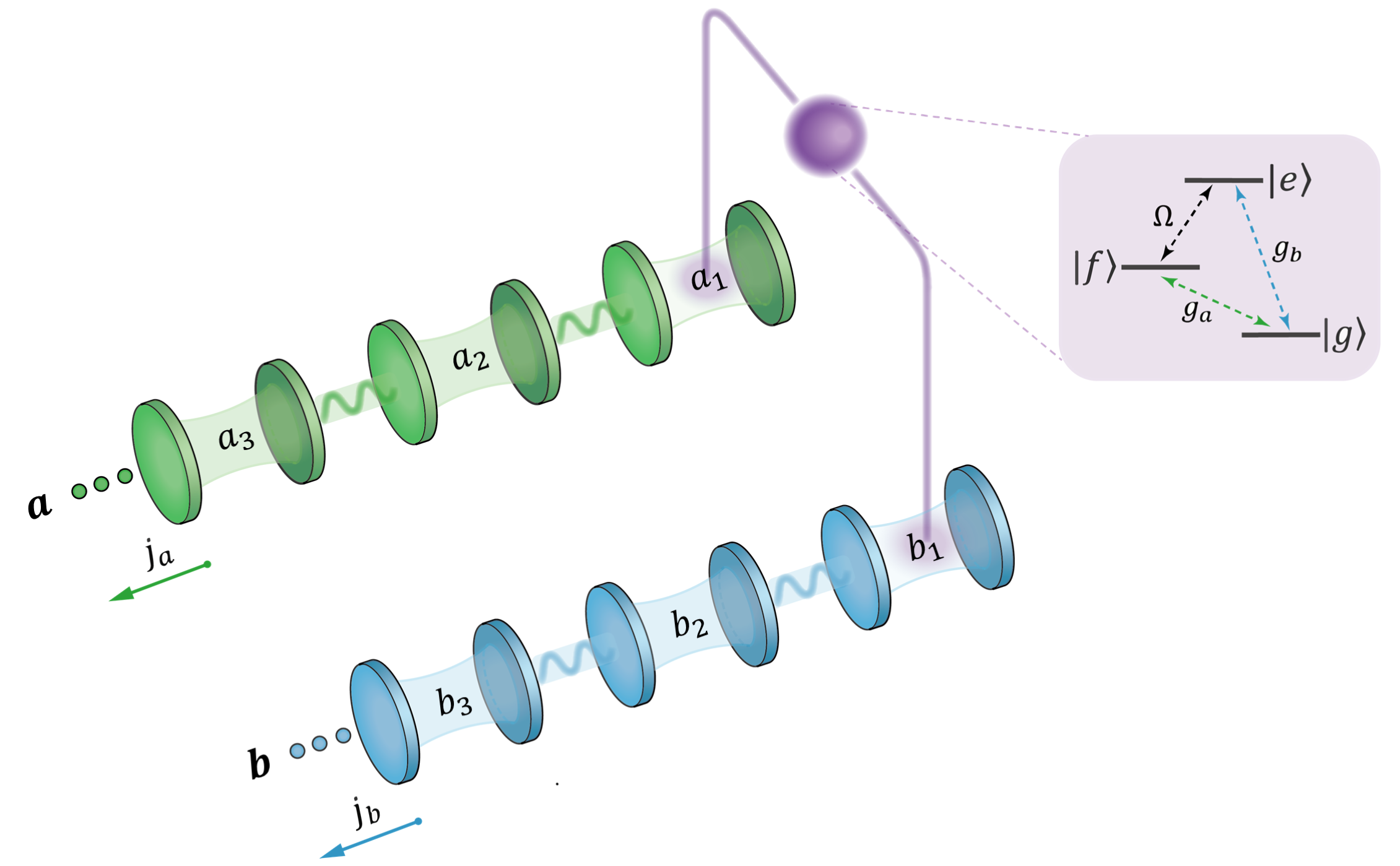}
	
	\caption{(Color online) Schematic configuration for a single-photon quantum router composed of a giant atom and two semi-infinite CRWs. The CRW denoted by green (blue) color is referred to as CRW-a (CRW-b) in the following context. The giant atom characterized by the ground state $|g\rangle$ and the two excited states $|f\rangle$ and $|e\rangle$. CRW-a (CRW-b) couples to the giant atom through the transition $|g\rangle\leftrightarrow|f\rangle$($|g\rangle\leftrightarrow|e\rangle$) with strength $g_a$($g_b$) and a classical field $\Omega$ is applied to resonantly drive $|f\rangle\leftrightarrow|e\rangle$ transition.}
	
	\label{fig1}
\end{figure*}

As depicted in Fig. \ref{fig1}, we consider a system consisting of a cyclic three-level giant atom and two 1D semi-infinite CRWs whose cavity modes are described by the annihilation operators $a_{j_a}$ and $b_{j_b}$, respectively, and subscript $j_d\left(d=a,b\right)=1,\cdots,+\infty$. The giant atom connects two CRWs through transitions $|g\rangle\leftrightarrow|f\rangle$ and $|g\rangle\leftrightarrow|e\rangle$ interacting with the cavity modes $a_1$ and $b_1$. The coupling strengths associated with these interactions are denoted as $g_a$ and $g_b$, respectively.
A classical field with freqency $\nu=\omega_e-\omega_f$ resonantly drives the transition $|f\rangle\leftrightarrow|e\rangle$ with Rabi freqency $\Omega$.

The total Hamiltonian of this system comprises three parts: the Hamiltonian of the two CRWs $\hat{H}_C$, the free Hamiltonian of the giant atom $\hat{H}_A$, and $\hat{H}_{CA}$ describes the interaction Hamiltonian among the giant atom, the classical field and the cavity modes, which can be expressed as $\hat{H}=\hat{H}_C+\hat{H}_A+\hat{H}_{CA}$, where ($\hbar =1$):
\begin{equation}\label{eq1}
	\hat{H}_C=\sum_{d=a, b} \sum_{j_d}\left[\omega_d \hat{d}_{j_d}^{\dagger} \hat{d}_{j_d}-\xi_d\left(\hat{d}_{j_d}^{\dagger} \hat{d}_{j_d+1}+\text { h.c. }\right)\right],
\end{equation}

\begin{equation}\label{eq2}
	\hat{H}_A=\omega_f |f\rangle\langle f|+\omega_e |e\rangle\langle e|,
\end{equation}

\begin{equation}\label{eq3}
	\hat{H}_{CA}=g_a |f\rangle\langle g| \hat{a}_1+g_b |e\rangle\langle g| \hat{b}_1+\Omega |e\rangle\langle f| e^{-i\nu t}+\text { h.c.},
\end{equation}
where $\omega_d\left(d=a,b\right)$ represents the same frequency of all cavity modes in CRW-d. $\xi_d$ is the hopping energies between any two nearest-neighbor cavities within CRW-d. $\omega_f$ ($\omega_e$) corresponds to the transition frequency between the giant atom's ground state $|g\rangle$ and the excited states $|f\rangle$ ($|e\rangle$).
In the rotating frame with respect to $\hat{H}_0=\omega_f\left(\sum_{j_a}\hat{a}_{j_a}^\dagger\hat{a}_{j_a}+|f\rangle\langle f|\right)+\omega_e\left(\sum_{j_b}\hat{b}_{j_b}^\dagger\hat{b}_{j_b}+|e\rangle\langle e|\right)$, the Hamiltonian of two CRWs described by typical tight-binding bosonic model\cite{crw_hamiltonian_1,crw_hamiltonian_2,crw_hamiltonian_3} transforms into:
\begin{equation}\label{eq4}
	\hat{H}_C'=\sum_{d=a, b} \sum_{j_d}\left[\Delta_d \hat{d}_{j_d}^{\dagger} \hat{d}_{j_d}-\xi_d\left(\hat{d}_{j_d}^{\dagger} \hat{d}_{j_d+1}+\text { h.c. }\right)\right],
\end{equation}
where $\Delta_{a\left(b\right)}=\omega_{a\left(b\right)}-\omega_{f\left(e\right)}$ represents the detuning between the cavity field frequency of CRW-a (CRW-b) and the atomic energy transition frequency $|g\rangle\leftrightarrow|f\rangle$ ($|g\rangle\leftrightarrow|e\rangle$). The interaction Hamiltonian between atomic energy transitions and the corresponding cavity modes becomes:
\begin{equation}\label{eq5}
	\hat{H}_{CA}'=g_a |f\rangle\langle g| \hat{a}_1+g_b |e\rangle\langle g| \hat{b}_1+\Omega |e\rangle\langle f|+\text { h.c.}.
\end{equation}

In this system, it's evident that the giant atom connects the originally independent two CRWs, and the switching of the classical field determines whether photons can undergo transfer within the two CRWs, acting as quantum channels for photon transmission.

\section{Single-Channel Quantum Router}\label{III}
It can be found that the operator
\begin{equation*}
	\hat{N}=\sum_{d=a,b}\sum_{j_d}\left(\hat{d}_{j_d}^\dagger \hat{d}_{j_d}\right)+|f\rangle\langle f|+|e\rangle\langle e|
\end{equation*}
commutes with the Hamiltonian $\hat{H}'=\hat{H}_0+\hat{H}_C'+\hat{H}_{CA}'$, and thus, the number of quanta is conserved in this system. In the single excitation scenario, incident photon either freely propagates in the two CRWs or is absorbed by the giant atom, occupying either the excited state $|f\rangle$ or, motivated by the classical field, the other excited state $|e\rangle$. This implies that the stationary eigenstate of this system can be expressed as:
\begin{equation}\label{eq6}
	\begin{split}
		|E\rangle =
		U_f|f, 0\rangle +U_e|e, 0\rangle+\sum_{j_a}A_{j_a}\hat{a}_{j_a}^{\dagger}|g, 0\rangle 
		+ \sum_{j_b}B_{j_b}\hat{b}_{j_b}^{\dagger}|g, 0\rangle,
	\end{split}
\end{equation}
where $|0\rangle$ is the vacuum state of the two CRWs. Here, $A_{j_a}$ and $B_{j_b}$ are the probability amplitudes of a single photon being in the $j_{a\left(b\right)}$th cavity of CRW-a and CRW-b, respectively. Additionally, $U_e$ and $U_f$ are the probability amplitudes for the three-level giant atom  in its corresponding excited states.
Substituting the Hamiltonian $\hat{H}_R=\hat{H}_C'+\hat{H}_{CA}'$ transformed through the rotating frame and the aforementioned eigenstates into the $Schr\ddot{o}dinger$ equation $\hat{H}_R|E\rangle =E|E\rangle$, then apply different left vectors to both sides of the equation, we can obtain a series of coupled stationary equations for all amplitudes:
\begin{equation}\label{eq7}
	\begin{split}
		&\Omega U_e+g_a A_1=E U_f,\\
		&\Omega U_f+g_b B_1=E U_e,\\
		&\delta_{j_a1} g_a U_f+\Delta_a A_{j_a}-\xi_a (A_{j_a-1}+A_{j_a+1})=E A_{j_a},\\
		&\delta_{j_b1} g_b U_e+\Delta_b B_{j_b}-\xi_b (B_{j_b-1}+B_{j_b+1})=E B_{j_b},
	\end{split}
\end{equation}
where $\delta_{mn}=1\left(0\right)$ for $m=n\left(m\not= n\right)$.
Eliminating the atomic amplitudes from Eq. \eqref{eq7} yields the discrete scattering equation for a single photon:
\begin{equation}\label{eq8}
	\begin{split}
		(\Delta_a-E) A_{j_a}=
		&\xi_a (A_{j_a-1}+A_{j_a+1})-\delta_{j_a1}(V_aA_1+GB_1),\\
		(\Delta_b-E) B_{j_b}=
		&\xi_b (B_{j_b-1}+B_{j_b+1})-\delta_{j_b1}(GA_1+V_bB_1).
	\end{split}
\end{equation}
Here, we introduce the energy-dependent $\delta$-like potential $V_d\equiv\frac{g_d^2 E}{E^2-\Omega^2}$, generated by the coupling of the giant atom with the CRWs. This potential, to some extent, reflects the scattering of photons when they encounter the giant atom in CRW-d. $G=\frac{g_a g_b \Omega}{E^2-\Omega^2}$ represents the effective dispersive coupling strength, influencing the probability of photon transfer between two CRWs.

The general form for the plane wave solution of Eq. \eqref{eq8} for a photon with energy $E$ incident from CRW-a, which then undergoes reflection and transfer when interacting with the giant atom\cite{t_shape_router_1}, can be written as:
\begin{equation}\label{eq9}
	\begin{split}	
		&A_{j_a}=
		\begin{cases}
			e^{-ik_aj_a}+r_1^ae^{ik_aj_a},&j_a>1\\
			A_1\sin{k_a},&j_a=1 
		\end{cases}\\
		&B_{j_b}=
		\begin{cases}
			t_1^be^{ik_bj_b},&\hspace{1.61cm}j_b>1\\
			B_1\sin{k_b},&\hspace{1.61cm}j_b=1 
		\end{cases}
	\end{split}
\end{equation}
where $r_1^a$ represents the reflection amplitude, $t_1^b$ denotes the amplitude of photon transfer from CRW-a to CRW-b, and $A_1$ ($B_1$) signifies the amplitude at the boundary cavity $j=1$ of CRW-a (CRW-b).
Substituting Eq. \eqref{eq9} into the discrete scattering equation Eq. \eqref{eq8}, in the region where $j_{a(b)}\not=1$, we can obtain the dispersion relationship of CRWs as $E_{d}=\Delta_{d}-2\xi_{d}\cos{k_{d}}$ ($k_{d} \in [0,2 \pi ]$), where the wave numbers $k_a$ and $k_b$ satisfy $\Delta_{a}-2\xi_{a}\cos{k_{a}}= \Delta_{b}-2\xi_{b}\cos{k_{b}}=E $. Conversely, in the region where $j_{a\left(b\right)}=1$, the reflection and transfer amplitude can be derived:
\begin{equation}\label{eq10}
	\begin{split}
		&r_1^a=-\frac{e^{ik}g_{a}^{2}g_{b}^{2}+\left(g_{a}^{2}+e^{2ik}g_{b}^{2}\right)E\xi+e^{ik}D}{e^{3ik}g_{a}^{2}g_{b}^{2}+e^{2ik}\left(g_{a}^{2}+g_{b}^{2}\right)E\xi+e^{ik}D},\\
		&t_1^b=\frac{Ng_{a}g_{b}\Omega\xi}{e^{3ik}g_{a}^{2}g_{b}^{2}+e^{2ik}\left(g_{a}^{2}+g_{b}^{2}\right)E\xi+e^{ik}D},
	\end{split}	
\end{equation}
here, $N=e^{2ik}-1$ represents the phase difference factor, and $D=\left(E^{2}-\Omega^{2}\right)\xi^{2}$ is a composite parameter involving the energy of the incident photon, classical field intensity, and hopping energies. To maximize the transfer efficiency, we set $\Delta_a=\Delta_b\equiv \Delta$ and $\xi_a=\xi_b\equiv \xi$, ensuring that both CRWs share identical structures. Thus, we always have  $k_a=k_b \equiv k$. It can be verified that the scattering amplitudes satisfy $|r_1^a|^2+|t_1^b|^2\equiv 1$, representing the conservation of probability flow for photon during the scattering process. The expression for $r_1^a$ in Eq. \eqref{eq10} indicates that $r_1^a\equiv -1$ always holds true when $k=n\pi$. This signifies photon will be reflected back to CRW-a with a probability of 100\%, a consequence of the quadratic dispersion close to the cosine band edge. This special case will not be considered in the subsequent discussions. Examining the expression for $t_1^b$, we find that the coupling strength $g_a$, $g_b$ and classical field $\Omega$, affect the probability of photon transfer to the other channel. Consequently toggling the classical field provides a convenient means to control the occurrence of photon transfer.

First, we discuss the scattering characteristics of single photon in this model based on the reflection and transfer amplitudes obtained above.

\begin{figure}[htb]
	
	\centering
	\includegraphics[width=12cm]{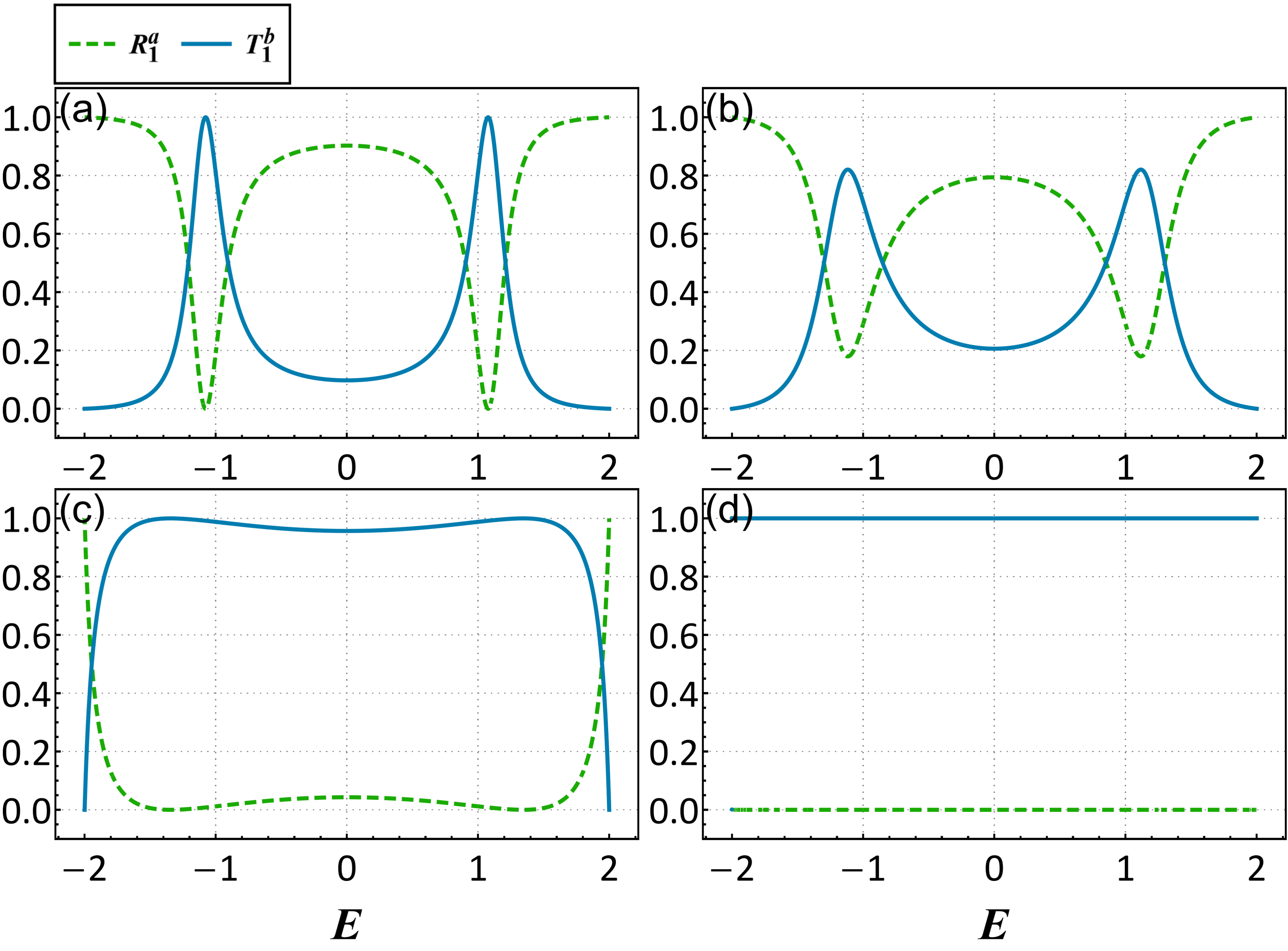}
	
	\caption{(Color online) The reflection rate (green dashed line) $R_1^a$ and the transfer rate (blue solid line) $T_1^b$ as functions of the energy of the incident photon.
		(a) $g_a=g_b=0.4$.
		(b) $g_a=0.4$, $g_b=0.6$.
		(c) $g_a=g_b=0.9$.
		(d) $g_a=g_b=1$.
		Here, we set $\Omega=1$. For convenience, all the parameters in this paper are in units of $\xi$, and we consistently set $\Delta=0$.}
	
	\label{fig2}
\end{figure}

In Fig. \ref{fig2}, we plot the reflection rate $R_1^a \equiv |r_1^a|^2$ and the transfer rate $T_1^b \equiv |t_1^b|^2$ as functions of the incident energy $E$ for various coupling strengths. 
First, we can observe that when the coupling strength is relatively weak, the peak of the transfer rate occurs near $E=\pm1$, as shown in Figs. \ref{fig2}(a) and \ref{fig2}(b). Part of the reason for this is consistent with what has been discussed in Ref.\cite{t_shape_router_1}, where a hard-wall boundary condition imposed on the radiation field by the termination of the CRW-a causes the excited state $|e\rangle$ of the atom to be dressed by its own radiation field with the Lamb shift of $\Delta(E)$\cite{t_shape_router_1}. 
However, the key difference lies in the presence of the classical field in this system, which also leads to an $\Omega$-dependent shift of the energy levels of $|e\rangle$\cite{x_shape_router_1,x_shape_router_2}. This is the reason behind the occurrence of two peaks in the transfer rate in this system, as opposed to the scenario presented in Ref.\cite{t_shape_router_1}, which does not involve a classical field and exhibits only one peak. Therefore, the peak of the transfer rate only appears when the incident energy $E$ corresponds to the energies of the two dressed states produced by the interaction of the atom with the classical field and its own radiation field.

Regarding the impact of coupling strengths, as shown in Figs. \ref{fig2}(a) and \ref{fig2}(b), it can be observed that when $g_a=g_b$, the peak of the transfer rate reaches its maximum. Moreover, from Fig. \ref{fig2}(b) to Fig. \ref{fig2}(c), it can be seen that as $g_ag_b$ increases, the peak broadens continuously which aligns with the observations found in Ref.\cite{x_shape_router_1,x_shape_router_2}. The unique aspect in our system is that the transfer rate's peak can reach 1, signifying perfect transfer. In comparison to Ref.\cite{t_shape_router_1}, which only has one transfer peak, our system exhibits two transfer peaks, implying that increasing the coupling strength can more effectively broaden the bandwidth of high transfer rates. As demonstrated in Fig. \ref{fig2}(c), when $g_a=g_b=0.9$, achieving over 90\% transfer rate within a large bandwidth ($-1.5<E<1.5$).

However, when we set $g_a=g_b=\Omega=1$ (or $g_a=g_b=\Omega=\xi$), as depicted in Fig. \ref{fig2}(d), an intriguing result emerges where the transfer rate remains constantly at 1, i.e., perfect transfer appears. In other words, this system can demonstrate frequency-independent scattering spectra, enabling our router to achieve desired routing results not just at specific frequencies, but across the entire energy band. This is a new and important result that other router schemes\cite{x_shape_router_1,t_shape_router_1,multi_coupled_points_crw_router} cannot achieve. It means that our router is no longer constrained by photon frequency, ensuring stability. Additionally, it greatly enhances the router's universality, as we can achieve the desired routing result with photons of any frequency within the energy band as information carriers.

\begin{figure}[htb]
	
	\centering
	\includegraphics[width=12cm]{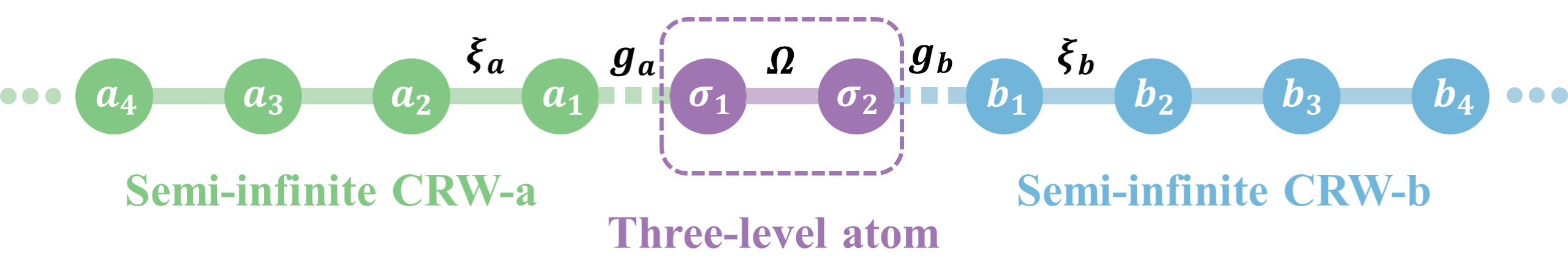}
	
	\caption{Equivalent schematic diagram of the system represented in Fig. \ref{fig1}, where three-level atom is interpreted as two emitters coupled by a classical field.}
	
	\label{fig3}
\end{figure}

To understand this phenomenon of perfect transfer, we now reinterpret the system depicted in Fig. \ref{fig1}. The three level atom can be regarded as two two-level emitters $\sigma_1$ and $\sigma_2$, coupled by classical field with strength $\Omega$, where $\sigma_1 \equiv |g\rangle\langle f|$, $\sigma_2 \equiv |g\rangle\langle e|$, and $\Omega |e\rangle\langle f|=\Omega \sigma_2^{\dagger} \sigma_1$ always holds (as defined in the Fig. \ref{fig3}). These emitters connect two semi-infinite CRWs through interaction between emitter $\sigma_1$ ($\sigma_2$) and $a_1$ ($a_2$), with coupling strength $g_a$ ($g_b$). Considering the conditions for perfect transfer, i.e., $g_a=g_b=\Omega=\xi_a=\xi_b$, in the single-excitation manifold, $\sigma_1$ and $\sigma_2$ behave like any other nodes in this system. Making the entire system equivalent to a uniformly complete infinite CRW. Thus, the transfer rate from semi-infinite CRW-a to semi-infinite CRW-b can be viewed as the transmission rate through equivalent infinite CRW, which is 1 for all frequencies.

\begin{figure}[htb]
	
	\centering
	\includegraphics[width=12cm]{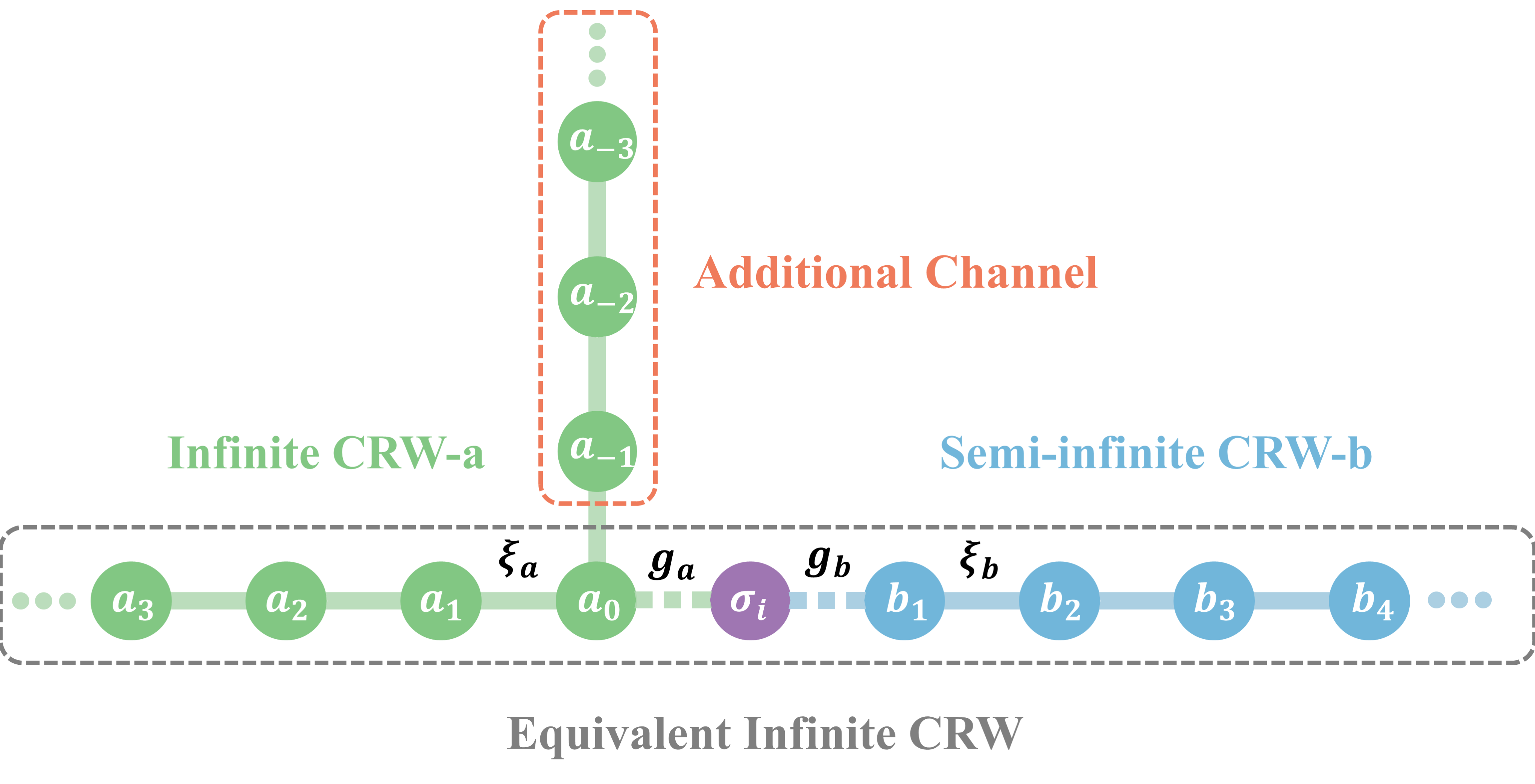}
	
	\caption{Equivalent schematic diagram of a router with an infinite CRW }
	
	\label{fig4}
\end{figure}

From this perspective, we can understand why no previous literature has achieved this intriguing result. In these prior schemes\cite{x_shape_router_1,t_shape_router_1,multi_coupled_points_crw_router}, at least one infinite CRW was used as a quantum channel. This results in an additional channel appearing, preventing the system from being equivalent to an infinite CRW, as shown in Fig. \ref{fig4}. Compared to the single-channel router depicted in Fig. \ref{fig1}, these schemes offer more output channels. However, by utilizing the multi-point coupling characteristics of giant atom to connect more semi-infinite CRWs, our multi-channel router (Fig. \ref{fig5}) achieves frequency-independent scattering spectra while also supporting multiple output channels, as detailed in Sec. \ref{IV}.

In addition to the intuitive physical explanation of perfect transfer mentioned above, we also provide a more rigorous analysis, revealing that it fundamentally originates from quantum interference effects between different pathways (see the Appendix \ref{appA} for more details).

\section{Multi-Channel Quantum Router}\label{IV}

In the previous sections, we explored the single-channel quantum router composed of a three-level giant atom and two semi-infinite CRWs. This router attains 100\% transfer rate across the entire energy band and can be controlled by the classical field to determine whether the transfer occurs. In this section, we will fully utilize this intriguing result to propose a multi-channel quantum router scheme that can achieve various functionalities while maintaining the router's stability and universality.

\begin{figure*}[htb]
	
	\centering
	\includegraphics[width=15cm]{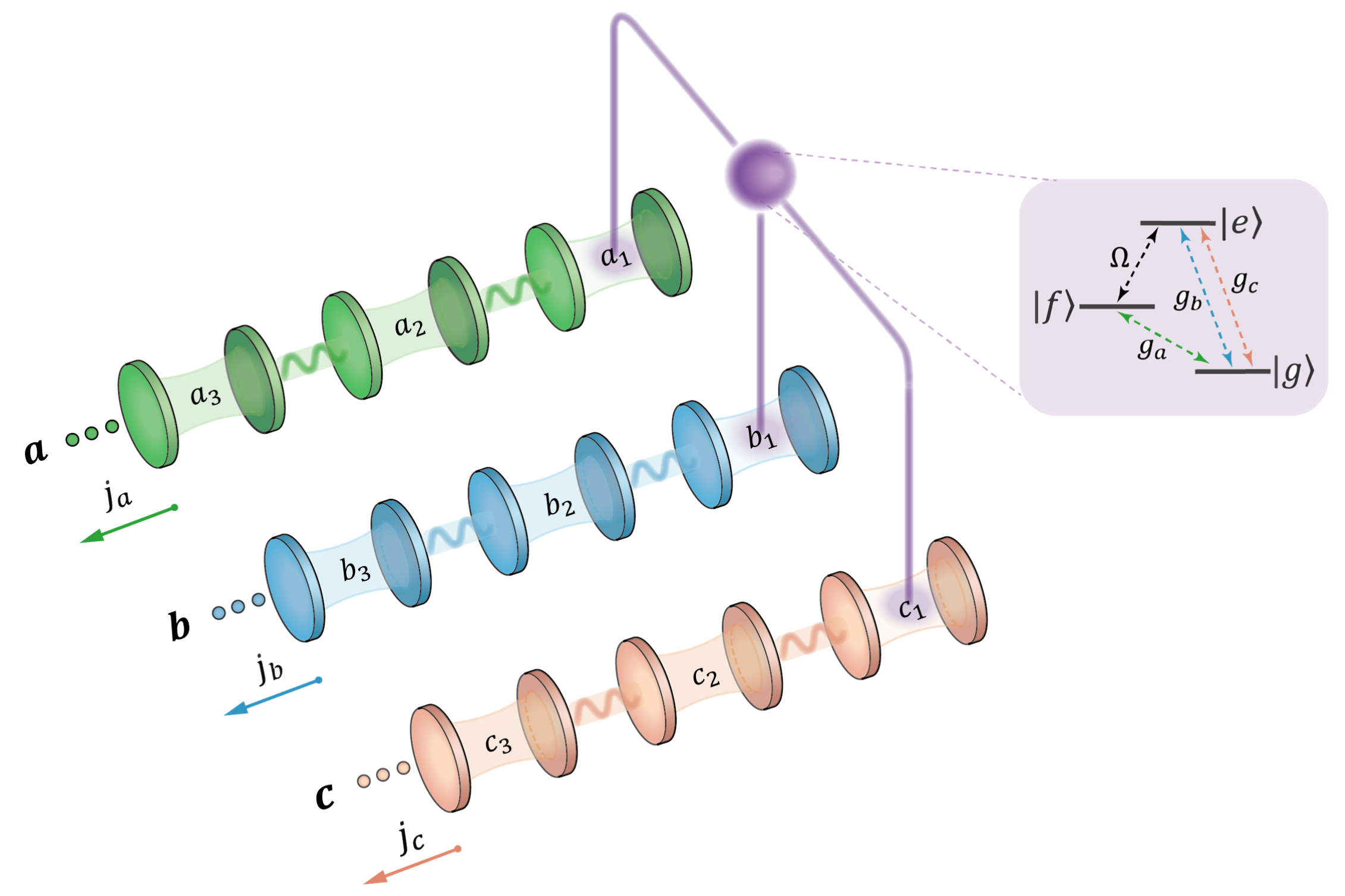}
	
	\caption{Schematic of a multi-channel quantum router comprises a giant atom and three semi-infinite CRWs, with one input channel and two output channels. The giant atom is designed to have three coupling points, each of which couples to the first cavity of three semi-infinite CRWs. Specifically, CRW-b (CRW-c) couples to the giant atom through the transition $|g\rangle\leftrightarrow|e\rangle$ with a strength of $g_b$ ($g_c$), while CRW-a couples to this atom through the transition $|g\rangle\leftrightarrow|f\rangle$ with a strength of $g_a$. Additionally, a classical field with a Rabi frequency of $\Omega$ is applied to drive the transition $|f\rangle\leftrightarrow|e\rangle$.}
	
	\label{fig5}
\end{figure*}

As shown in Fig. \ref{fig5}, we additionally introduce a semi-infinite CRW-c characterized by frequency $\omega_c$ and hopping strength $\xi_c$ to the model represented in Fig. \ref{fig1}. In this case, the system's Hamiltonian is given by:
\begin{equation}\label{eq15}
	\begin{split}
		&\hat{H}_C=\sum_{d=a,b,c} \sum_{j_d}\left[\omega_d \hat{d}_{j_d}^{\dagger} \hat{d}_{j_d}-\xi_d\left(\hat{d}_{j_d}^{\dagger} \hat{d}_{j_d+1}+\text { h.c. }\right)\right],\\
		&\hat{H}_A=\omega_f |f\rangle\langle f|+\omega_e |e\rangle\langle e|,\\
		&\hat{H}_I=g_a |f\rangle\langle g| \hat{a}_1+g_b |e\rangle\langle g| \hat{b}_1+g_c |e\rangle\langle g| \hat{c}_1+\Omega |e\rangle\langle f| e^{-i\nu t}+\text { h.c.}.
	\end{split}
\end{equation}
Following a similar process as outlined above, with some algebraic manipulations, we can readily derive the discrete scattering equation for a single photon in this model:
\begin{equation}\label{eq16}
	\begin{split}
		&\Delta_a A_{j_a}-\xi_a (A_{j_a-1}+A_{j_a+1})+\delta_{j_a1}(V_aA_1+G_{ab}B_1+G_{ac}C_1)=EA_{j_a},\\
		&\Delta_b B_{j_b}-\xi_b (B_{j_b-1}+B_{j_b+1})+\delta_{j_b1}(G_{ab}A_1+V_bB_1+G_{bc}C_1)=EB_{j_b},\\
		&\Delta_c C_{j_c}-\xi_c (C_{j_c-1}+C_{j_c+1})+\delta_{j_c1}(G_{ac}A_1+G_{bc}B_1+V_cC_1)=EC_{j_c}.
	\end{split}	
\end{equation}
The class-$\delta$ potentials, here denoted as $V_d=\frac{g_d^2 E}{E^2-\Omega^2}$ (for $d=a,b,c$), are consistent with the single-channel quantum router depicted in Fig. \ref{fig1}. Furthermore, the effective dispersive coupling strengths between CRW-a and b ($G_{ab}=\frac{\Omega g_ag_b}{E^2-\Omega^2}$), CRW-a and c ($G_{ac}=\frac{\Omega g_ag_c}{E^2-\Omega^2}$), and CRW-b and c ($G_{bc}=\frac{Eg_bg_c}{E^2-\Omega^2}$) determine the extent of coupling interactions. It is evident that the classical field switch has the exclusive power to dictate the presence of coupling interactions between CRW-a and b and between CRW-a and c, while it does not impact the coupling between CRW-b and c.

When a photon is incident from CRW-a in the form of a plane wave, The probability amplitudes in the asymptotic regions are given by
\begin{equation}\label{eq17}
	\begin{split}
		&A_{j_a}=
		\begin{cases}
			e^{-ik_aj_a}+r_3^ae^{ik_aj_a},&j_a>1\\
			A_3\sin{k_a},&j_a=1 
		\end{cases}\\
		&B_{j_b}=
		\begin{cases}
			t_3^be^{ik_bj_b},&\hspace{1.61cm}j_b>1\\
			B_3\sin{k_b},&\hspace{1.61cm}j_b=1 
		\end{cases}\\
		&C_{j_c}=
		\begin{cases}
			t_3^ce^{ik_cj_c},&\hspace{1.61cm}j_c>1\\
			C_3\sin{k_c},&\hspace{1.61cm}j_c=1 
		\end{cases}
	\end{split}	
\end{equation}
Here, $r_3^a$ and $t_3^b$ ($t_3^c$) represent the reflection and transfer amplitudes of a photon after scattering by the giant atom, resulting in reflection or transmission to CRW-b (CRW-c). $A_3, B_3, C_3$ respectively denote the amplitudes at the boundary cavity $j_d=1$ of CRW-d. Substituting Eq. \eqref{eq17} into the discrete scattering equation Eq. \eqref{eq16} and setting $j_a=1, j_b=1, j_c=1$, we can subsequently determine the amplitudes for reflection and transfer:
\begin{equation}\label{eq18}
	\begin{split}
		&r_3^a=-\frac{e^{ik}g_{a}^{2}\left(g_{b}^{2}+g_{c}^{2}\right)+\left[g_{a}^{2}+e^{2ik}\left(g_{b}^{2}+g_{c}^{2}\right)\right]E\xi-e^{ik}D}{e^{3ik}g_{a}^{2}\left(g_{b}^{2}+g_{c}^{2}\right)+e^{2ik}\left(g_{a}^{2}+g_{b}^{2}+g_{c}^{2}\right)E\xi+e^{ik}D},\\
		&t_3^b=\frac{Ng_{a}g_{b}\Omega\xi}{e^{3ik}g_{a}^{2}\left(g_{b}^{2}+g_{c}^{2}\right)+e^{2ik}\left(g_{a}^{2}+g_{b}^{2}+g_{c}^{2}\right)E\xi+e^{ik}D},\\
		&t_3^c=\frac{Ng_{a}g_{c}\Omega\xi}{e^{3ik}g_{a}^{2}\left(g_{b}^{2}+g_{c}^{2}\right)+e^{2ik}\left(g_{a}^{2}+g_{b}^{2}+g_{c}^{2}\right)E\xi+e^{ik}D},
	\end{split}	
\end{equation}
where, we similarly constrain three CRWs to have identical structures to achieve optimal transfer efficiency and the flow conservation relation changes into $|r_3^a|^2+|t_3^b|^2+|t_3^c|^2\equiv 1$. It can be observed in Eq. \eqref{eq18} that the expression for scattering amplitude shares a similar overall form with the single-channel scenario discussed in the Sec.\ref{III}. In both cases, the denominators consist of terms involving the product and sum of different coupling strength squares, along with the composite parameter $D$. Additionally, the phase factors for each term remain consistent in the both situations. Examining the expressions for $t_3^b$ and $t_3^c$ in Eq. \eqref{eq18}, we find that the coupling strength $g_b$ or $g_c$ is the sole parameter to distinguish the two transfer amplitudes. Therefore, effective control over the proportion of photon transfer to different channels can be achieved by adjusting the relative strengths of $g_a$ and $g_b$. Meanwhile, toggling of the classical field serves the purpose of controlling whether the transfer occurs. Next, we will explore the functionalities of this model based on these scattering amplitudes. Firstly, one of its achievable functionality is the 100\% probability control of photon output from any channel.

\begin{figure}[htb]
	
	\centering
	\includegraphics[width=12cm]{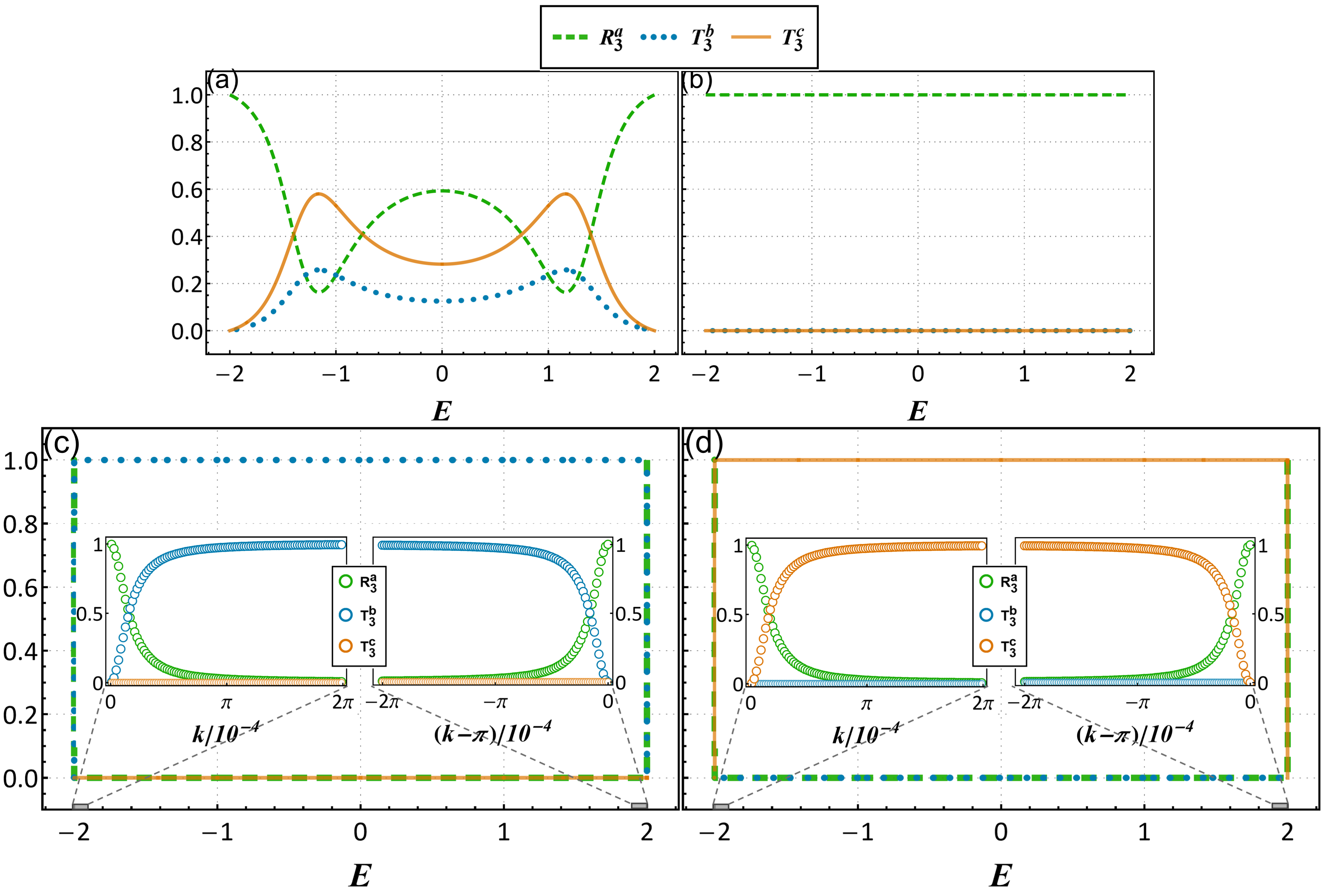}
	
	\caption{(color online) Reflection rate (green dashed line) $R_3^a$, the transfer rate to CRW-b (blue dotted line) $T_3^b$, and the transfer rate to CRW-c (orange solid line) $T_3^c$, are plotted as functions of the incident photon energy $E$ for the multi-channel quantum router, with different parameter settings:
		(a) $g_a=0.5$, $g_b=0.4$, $g_c=0.6$, $\Omega=1$.
		(b) $\Omega=0$.
		(c) $g_a=g_b=\Omega=1$, $g_c=0.01$.
		(d) $g_a=g_c=\Omega=1$, $g_b=0.01$.
		The insets present, in a scatter plot style, detailed variations of the scattering spectra at the band edges, where green circles represent $R_3^a$, blue circles represent $T_3^b$, and orange circles represent $T_3^c$.
	}
	
	\label{fig6}
\end{figure}

In Fig. \ref{fig6}, we present the output scenarios of this model under various parameter settings. As shown in Fig. \ref{fig6}(a), when the coupling strengths between the giant atom and different CRWs are relatively similar, and there is the presence of a classical field, photon incident from CRW-a will be distributed among three channels in certain proportions (which depends on the ratio of the incident energy to the coupling strengths). However, when there is no classical field, as depicted in Fig. \ref{fig6}(b), CRW-a decouples from the other two CRWs, and photon incident from CRW-a will be 100\% likely to output from channel a. By adjusting three coupling strengths to satisfy $g_a=g_b=\Omega=1\gg g_c$, as shown in Fig. \ref{fig6}(c), photon incident from CRW-a can be almost entirely transferred to channel b (in fact, setting $g_c=0$ directly achieves 100\% transfer rate). And furthermore, we observe that around $E=\pm 2$, $R_3^a$ and $T_3^b$ behave like two vertical lines. To explain this phenomenon, we use a scatter plot style in the insets of Fig. \ref{fig6}(c) to show the scattering spectra at the band edges, i.e., when the incident photon wave number $k\in\left[0,2\pi\times10^{-4}\right]$ for the left plot and $k\in\left[9998\pi\times10^{-4},\pi\right]$ for the right plot. Each point in those plots corresponds to a sampling interval of $\delta k=2\pi\times10^{-6}$.
From those insets, we note that around $E=-2$ (or $k=0$), the distribution of points on the curves of $R_3^a$ and $T_3^b$ transitions rapidly from sparse to dense, indicating that they decrease from 1 to 0 and increase from 0 to 1 with a significant gradient as the energy increases, respectively. Conversely, around $E=2$ (or $k=\pi$), they exhibit the opposite trend. Therefore, the phenomenon of two vertical curves appearing at the band edges is attributed to the fact that $R_3^a$ and $T_3^b$ undergo nearly all their variations within a very small momentum range ($k\in\left[0,2\pi\times10^{-4}\right]$). Physically, this is reasonable because at the band edges, i.e. $k=0$ or $k=\pi$, corresponding to quadratic dispersion of photon, no transfer can occur. Once the momentum of the incident photon skips these two values, their reflection amplitudes rapidly decrease to 0 due to interference effects, and correspondingly, transfer amplitudes increase rapidly to 1.
From the insets in Fig. \ref{fig6}(c), we also observe that $T_3^c$ is nearly 0 across the entire band. This is because the coupling strength between the giant atom and CRW-c is much smaller than that with CRW-b, causing almost all the photon transferred from CRW-a to flow into CRW-b.
Similarly, by adjusting three coupling strengths to satisfy $g_a=g_c=\Omega=1\gg g_b$, as shown in Fig. \ref{fig6}(d), photon incident from CRW-a can be almost entirely transferred to channel c. Regarding the scattering at the band edges, it is analogous to the scenario depicted in Fig. \ref{fig6}(c), however here, photon transferred from CRW-a almost entirely flows into CRW-c.
Thus, the results in Fig. \ref{fig6} imply that by manipulating the strengths of three couplings and the  classical field, one can freely control the incident photons to be output from any channel with a 100\% transfer rate.

\begin{figure}[htb]
	
	\centering
	\includegraphics[width=12cm]{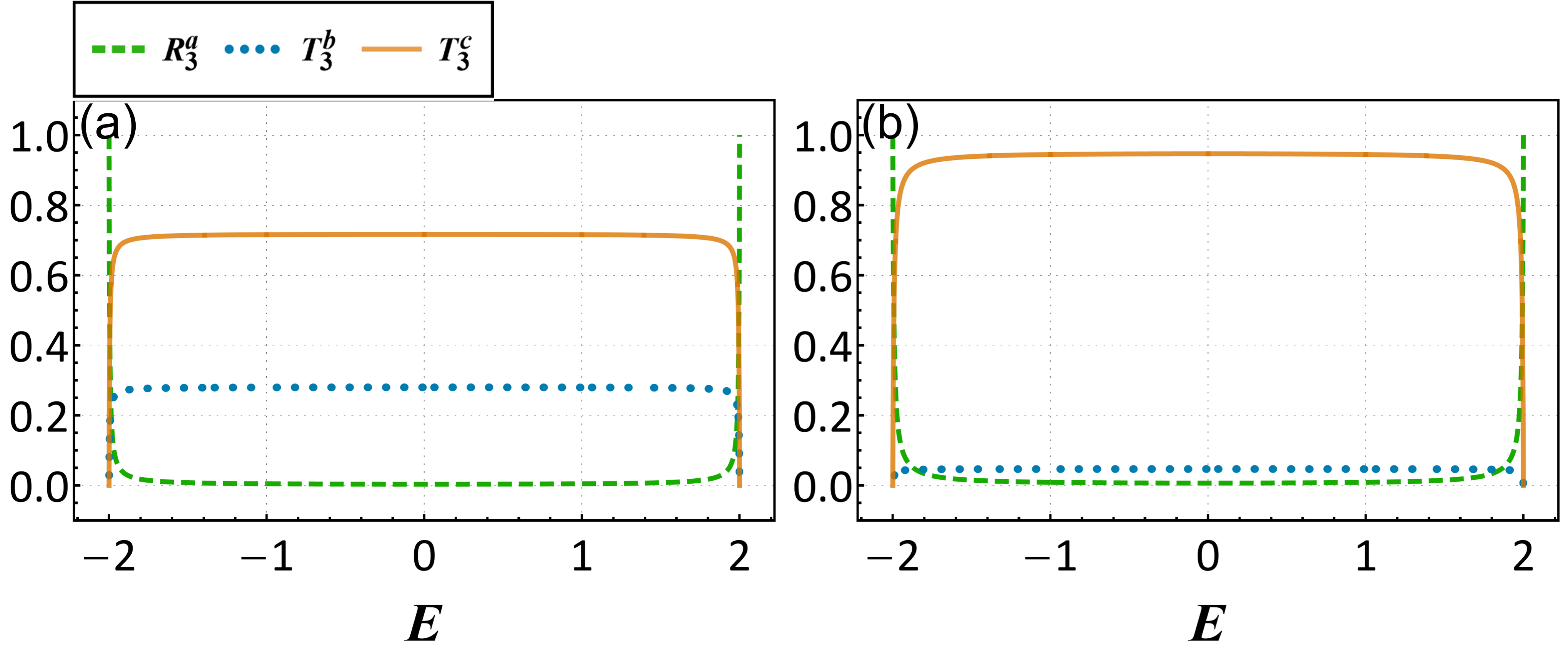}
	
	\caption{(color online) Reflection coefficient (green dashed line) $R_3^a$, the transfer coefficient to CRW-b (blue dotted line) $T_3^b$, and the transfer coefficient to CRW-c (orange solid line) $T_3^c$, as functions of the incident photon energy $E$, under different parameter configurations for the multi-channel quantum router.
		(a) $g_b=0.5,g_c=0.8$;
		(b) $g_b=0.2,g_c=0.9$. 
		Here, we have set $g_a=\Omega=1$.}
	
	\label{fig7}
\end{figure}

However, in quantum networks, quantum nodes sometimes need to not only control the output of photons carrying information from a single channel but also flexibly control the output of photons from two or more channels in the desired proportion. Such as quantum multiplexing applied to quantum error correction\cite{quantum_network_4} and distributed quantum computing\cite{quantum_computing_1}. 
In this cases, we only need to appropriately adjust the parameters mentioned above to achieve stable output from channels b and c in the desired ratio within the interval $0<g_b, g_c\le1$, while satisfying $g_a=\Omega=1$. As shown in Fig. \ref{fig7}, by properly adjusting the coupling strengths $g_a$ and $g_b$ between the giant atom and the two output channels, we can simultaneously output photon from two channels in the desired proportion. Moreover, the ratio of output $T_3^b/T_3^c$ is consistently proportional to $g_b^2/g_c^2$, which can be easily derived from Eq. \eqref{eq18}. Due to the setting of $g_a=\Omega=1$, the total probability of photon output from the two channels is almost 1, and the output ratio remains nearly constant across the entire energy band, ensuring the device's high fidelity and stability.

In summary, under the collaborative influence of classical fields and various couplings, the multi-channel quantum router can achieve: (i) stable lossless output of photon carrying information from any port, and (ii) simultaneous control of photon to output from two ports in the desired proportions. It is worth noting that while other routers\cite{x_shape_router_1,t_shape_router_1,multi_coupled_points_crw_router} can also achieve these functionalities, their scattering spectra are highly sensitive to the frequency of the incident photon. This sensitivity necessitates strict control of photon frequencies to achieve the desired routing effect. Our scheme, however, cleverly avoids this difficulty and significantly enhances the router's universality, as we can use photons of any frequency within the energy band as information carriers. Theoretically, with the assistance of giant atoms, this model can be further extended to accommodate any number of input and output channels by connecting all semi-infinite CRWs serving as input channels to the same transition of the cyclic three-level giant atom, while coupling CRWs acting as output channels to another transition. In such a scenario, the classical field can regulate the flow of information from the output ports, and diverse input-output configurations for different ports can be flexibly managed by adjusting the coupling strengths between giant atoms and various CRWs.

\begin{figure}[htb]
	
	\centering
	\includegraphics[width=15cm]{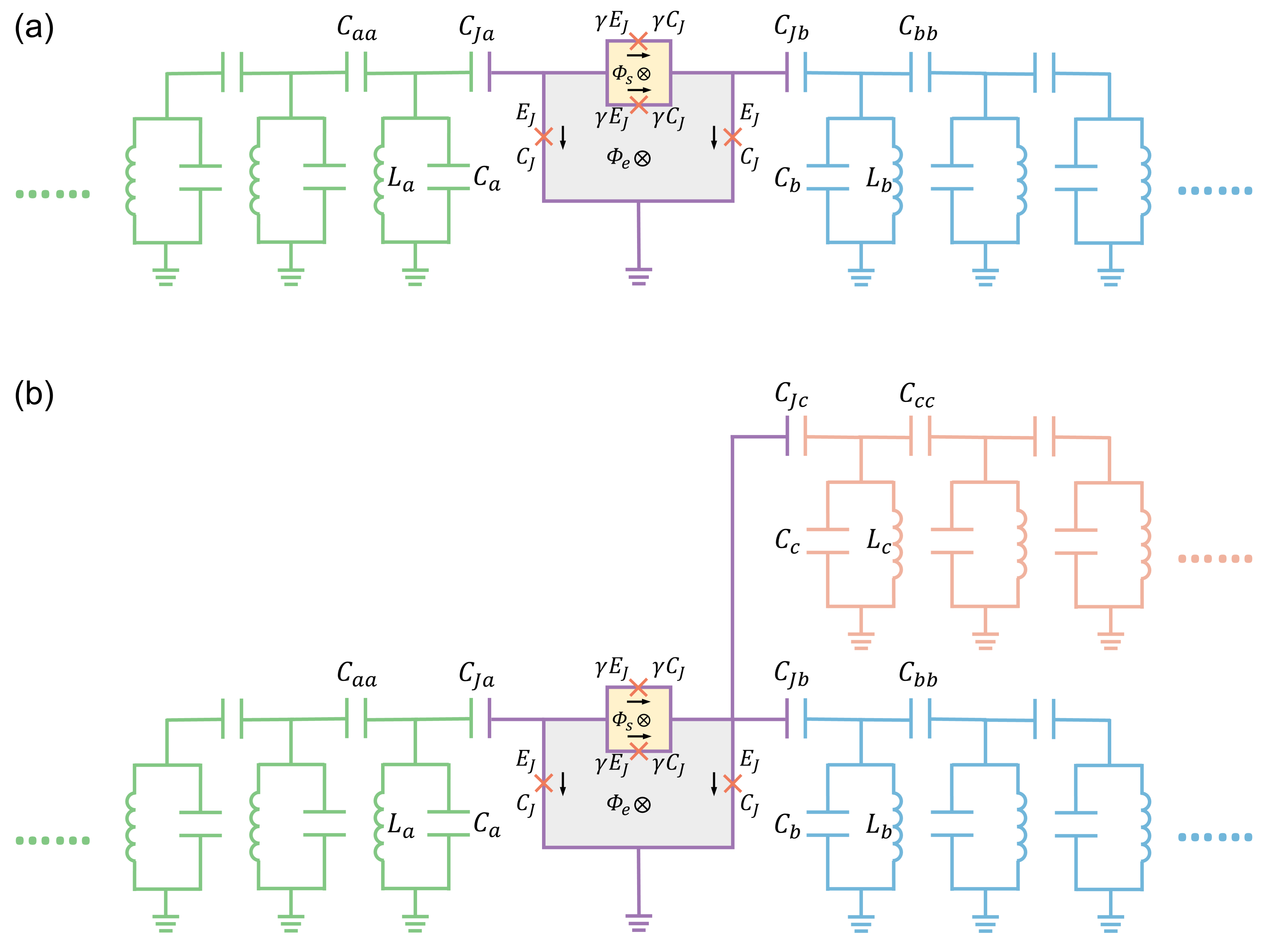}
	
	\caption{(color online) Schematic diagram for the possible circuit to implement the router model based on superconducting quantum circuits, (a) for the single-channel router in Fig. \ref{fig1}, and (b) for the multi-channel router in Fig. \ref{fig5}. The green LCCs-chain consists of LC circuits with capacitance $C_a$ and inductance $L_a$, where adjacent LC circuits are coupled through coupling capacitors $C_aa$, corresponding to semi-infinite CRW-a. Similarly, the blue (orange) LCCs-chain corresponds to semi-infinite CRW-b (c). The superconducting artificial atom (purple) is structured by a symmetric SQUID and two identical Josephson junctions with coupling energy $E_J$ and capacitance $C_J$, corresponding to the cyclic three-level giant atom. In the symmetric SQUID, the other two Josephson junctions have coupling energy $\gamma E_J$ and capacitance $\gamma C_J$. Two different loops are pierced by externally applied magnetic fluxes $\Phi_e$ (grey) and $\Phi_s$ (yellow), respectively. The superconducting artificial atom is coupled to different LCCs chains simultaneously through capacitor $C_{J\alpha}$.}
	
	\label{fig8}
\end{figure}

In the final part of this section, we briefly discuss potential experimental setups. We believe that superconducting quantum circuits are the most promising platform for realizing this system\cite{SQC_1,SQC_2,SQC_3}. As shown in Fig. \ref{fig8}, the semi-infinite CRWs in Figs. \ref{fig1} and \ref{fig5} are obtained by LC circuits-chains (LCCs-chains)\cite{SQC_CRW_3,SQC_4}. 
The LCCs-chain corresponding to semi-infinite CRW-$\alpha$ ($\alpha=a,b,c$) consists of LC circuits with capacitance $C_{\alpha}$ and inductance $L_{\alpha}$, where adjacent LC circuits are coupled through coupling capacitors $C_{\alpha \alpha}$.
The cyclic three-level atom is structured by a symmetric superconducting quantum interference device (SQUID) and two identical Josephson junctions with coupling energy $E_J$ and capacitance $C_J$\cite{SQC_GA_1,Selection_Rule}. In the symmetric SQUID, the other two Josephson junctions have coupling energy $\gamma E_J$ and capacitance $\gamma C_J$. The superconducting loop containing Josephson junction $E_J$ is pierced by externally applied magnetic flux (grey) $\Phi_e$, while the applied flux (yellow) threading through the symmetric SQUID is $\Phi_s$. This superconducting artificial atom is coupled to CRW-a, b, and c via capacitors $C_{Ja}$, $C_{Jb}$, and $C_{Jc}$\cite{SQC_CRW_2,SQC_GA_2}, respectively, enabling the multi-point coupling functionality of the giant atom. In fact, the concept of a giant atom is primarily used to address the coupling issues between superconducting qubits (artificial atoms) and surface acoustic waves (microwave photons) in superconducting quantum circuit platforms\cite{giant_atom_1,giant_atom_2,giant_atom_engineering}. We believe that the proposed scheme is experimentally feasible and deserves to be tested under the currently existing experimental technique.

\section{Conclusion}\label{V}

In this study, we first propose a theoretical model for single-channel quantum router, which connects two semi-infinite CRWs through a giant atom. It is demonstrated that the scattering spectra of this model exhibit two transfer peaks with a maximum value of 1 due to the influence of the classical field. Unlike other routers with only one transfer peak, the enhancement of the giant atom-cavity coupling strength can be proved to be a more efficient means of broadening the energy range with high transfer rates. More importantly, this router can achieve 100\% transfer rate over the entire energy band and can be controlled by the classical field to determine whether the transfer occurs. We have provided an intuitive physical explanation for this intriguing result, and verified that it fundamentally originates from quantum interference effects between different pathways.

Furthermore, we expand the capabilities of the single-channel quantum router by incorporating an additional semi-infinite CRW. The results demonstrate that by adjusting the strengths of different couplings and the classical field, We can control the output of photons from any single port with perfect transfer or distribute them simultaneously to both ports in any desired proportion. More importantly, the scattering spectra is almost unaffected by the photon frequency, allowing us to avoid frequency constraints and significantly enhance the router's stability and universality. Moreover, by coupling giant atoms with more semi-infinite CRWs, this router can achieve greater flexibility and complexity in its applications.

\section{ACKNOWLEDGMENTS}
This work was supported by National Natural Science Foundation of China (Grants No. 11874190 and No. 12247101). Support was also provided by Supercomputing Center of Lanzhou University.

\appendix
\section{The Underlying Mechanism of Perfect Transfer}\label{appA}

In this Appendix, we provide a more rigorous analysis of perfect transfer, revealing that this intriguing result fundamentally originates from quantum interference effects between different pathways.

In the quantum router represented by Fig. \ref{fig1}, a photon incident in the form of a plane wave from CRW-a initially propagates freely to the right along CRW-a. When it reaches cavity $a_1$, it is either captured by the giant atom or reflected by the boundary. The portion reflected by the boundary propagates along the positive $j_a$ axis, accompanied by a phase shift of $\pi$, and becomes part of the final reflected wave. The portion captured by the giant atom is subsequently re-emitted into two CRWs, with the wave propagating along positive $j_a$ axis forming another component of the final reflected wave. Notably, the process of the giant atom absorbing and re-emitting the photon does not alter the photon's phase.
In reality, the final reflected wave is a superposition of the two types of waves described above. Due to the hard boundary conditions, these two waves always have a phase difference of $\pi$, so under the appropriate parameter conditions, destructive interference occurs between them\cite{destructive_interference_1,destructive_interference_2,destructive_interference_3,destructive_interference_4}, causing the amplitude of the reflected wave to be 0 and so the amplitude of the transferred wave to be 1.

Based on the qualitative analysis mentioned above, this phenomenon of perfect transfer arises from the phase difference between two pathways forming the final reflected wave. Thus, we simply need to manipulate the phase of one of the pathways, for example, by altering the coupling position of the giant atom with CRW-a. This will ensure that two waves contributing to the final reflection wave are not reflected or re-emitted at the same position $j_a$. One can validate this result by investigating the following model.

\begin{figure}[htb]
	\centering
	\includegraphics[width=15cm]{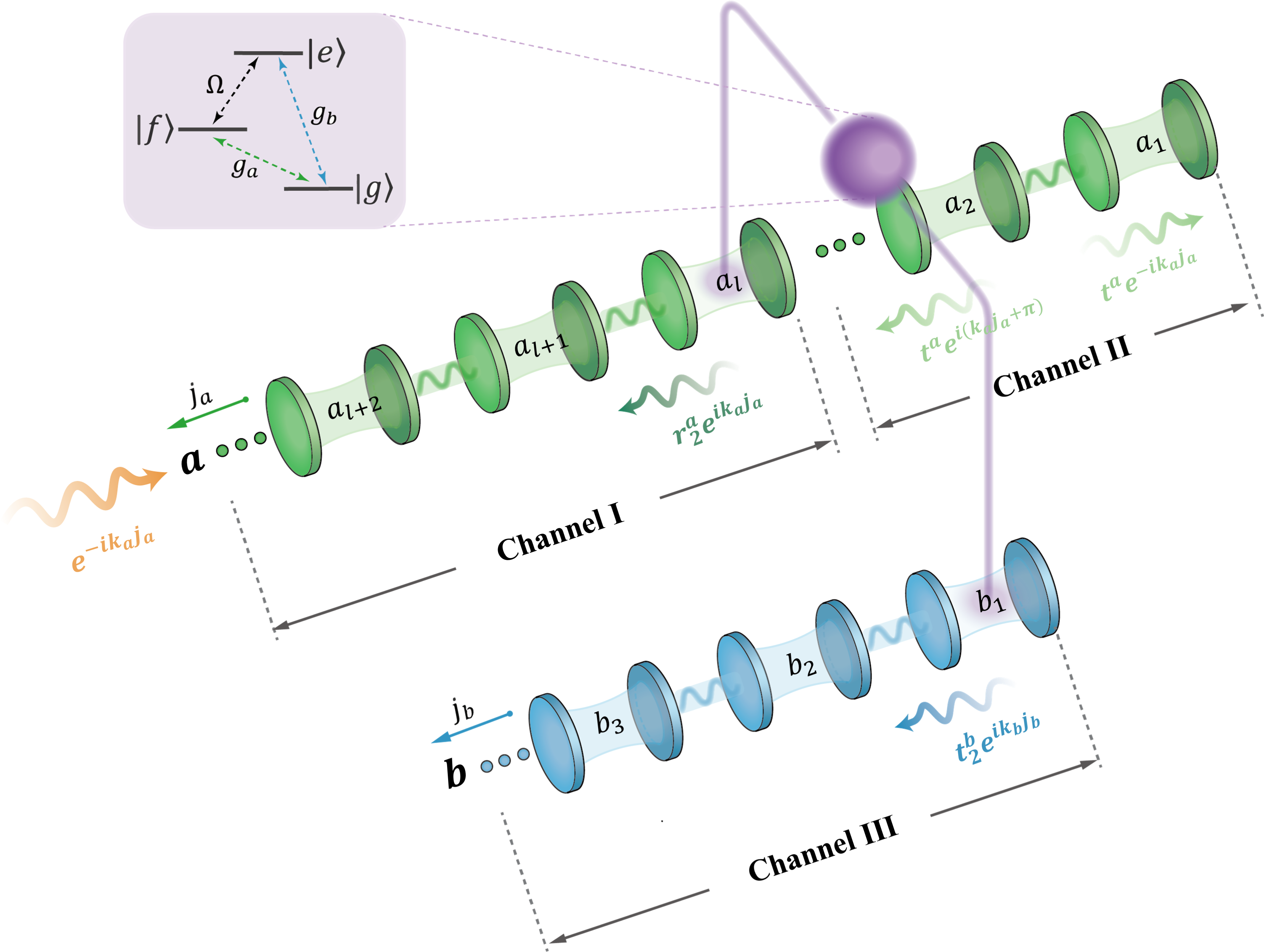}
	
	\caption{Schematic diagram of single-photon scattering in adjusted single-channel quantum router. The router also comprises a giant atom and two semi-infinite CRWs, with the two coupling
		points of the giant atom couple to lth cavity of CRW-a and the first cavity of CRW-b, respectively. The photon is incident on CRW-a in the form of a plane wave, and undergoes scattering upon encountering the scattering cavity. Channel I denotes the cavities in CRW-a with $j_a \geq l$, where a final reflected wave $r_2^a e^{ik_aj_a}$ exists in steady state. Channel II represents cavities in CRW-a with $1 \leq j_a < l$, where a standing wave, formed by the superposition of two plane waves with equal amplitude, opposite direction, and a phase difference of $\pi$, is established in steady state. Channel III represents CRW-b, where a transfer wave $t_2^b e^{ik_bj_b}$ exists in steady state.}
	
	\label{figs2}
\end{figure}

As depicted in Fig. \ref{figs2}, we change the coupling position of the giant atom with CRW-a, specifically, coupling the giant atom with $l$th cavity of CRW-a. Following the same steps as outlined in Sec. \ref{II} and Sec. \ref{III}, we can readily derive the discrete scattering equation for a single photon in this model.
\begin{equation}\label{eq11}
	\begin{split}
		\Delta_a A_{j_a}-\xi_a (A_{j_a-1}+A_{j_a+1})+\delta_{j_al}(V_aA_l+GB_1)
		&=EA_{j_a},\\
		\Delta_b B_{j_b}-\xi_b (B_{j_b-1}+B_{j_b+1})+\delta_{j_b1}(GA_l+V_bB_1)
		&=EB_{j_b}.
	\end{split}	
\end{equation}

When a photon in the form of a plane wave is incident from CRW-a, the solution to Eq. \eqref{eq11} can be expressed in the following form \cite{t_shape_router_1}:
\begin{equation}\label{eq12}
	\begin{split}
		&A_{j_a}=
		\begin{cases}
			e^{-ik_aj_a}+r_2^ae^{ik_aj_a},&j_a>l\\
			A_2\sin{k_aj_a},&1\le j_a<l 
		\end{cases}\\
		&B_{j_b}=
		\begin{cases}
			t_2^be^{ik_bj_b},&\hspace{1.61cm}j_b>1\\
			B_2\sin{k_b},&\hspace{1.61cm}j_b=1 
		\end{cases}
	\end{split}	
\end{equation}
here, $r_2^a$ and $t_2^b$ represent the reflection and transfer amplitudes of a photon after scattering in $a_l$ cavity. $A_2$ represents the amplitude of the standing wave in the region of CRW-a with $1\le j_a<l$, and $B_2$ denotes the amplitude at the boundary cavity of CRW-b with $j=1$. By substituting Eq. \eqref{eq12} into the discrete scattering equation Eq. \eqref{eq11} and setting $j_a=l$ and $j_b=1$, one can get the reflection and transfer amplitudes:
\begin{equation}\label{eq13}
	\begin{split}
		&r_2^a=\frac{e^{2ik}M_{-}g_{a}^{2}g_{b}^{2}+e^{ik}\left(M_{-}g_{a}^{2}-Ng_{b}^{2}\right)E\xi-ND}{e^{2ik}M_{+}g_{a}^{2}g_{b}^{2}+e^{ik}\left(M_{+}g_{a}^{2}+Ng_{b}^{2}\right)E\xi+ND},\\
		&t_2^b=\frac{e^{-ikl}M_{+}Ng_{a}g_{b}\Omega\xi}{e^{2ik}M_{+}g_{a}^{2}g_{b}^{2}+e^{ik}\left(M_{+}g_{a}^{2}+Ng_{b}^{2}\right)E\xi+ND},
	\end{split}	
\end{equation}   
similarly, we assume that those two CRWs have identical structures. Here, $M_{\pm}=e^{\pm2ikl}-1$ characterizes the phase variation resulting from changes in the coupling position between the giant atom and CRW-a, while $N=e^{2ik}-1$ and $D=\left(E^{2}-\Omega^{2}\right)\xi^{2}$ have already been introduced in Sec. \ref{III}. It is evident from Eq. \eqref{eq13} that when $l=1$, corresponding to the coupling of the giant atom with the first cavity of CRW-a, it simplifies to Eq. \ref{eq10}.

\begin{figure}[htb]
	\centering
	\includegraphics[width=12cm]{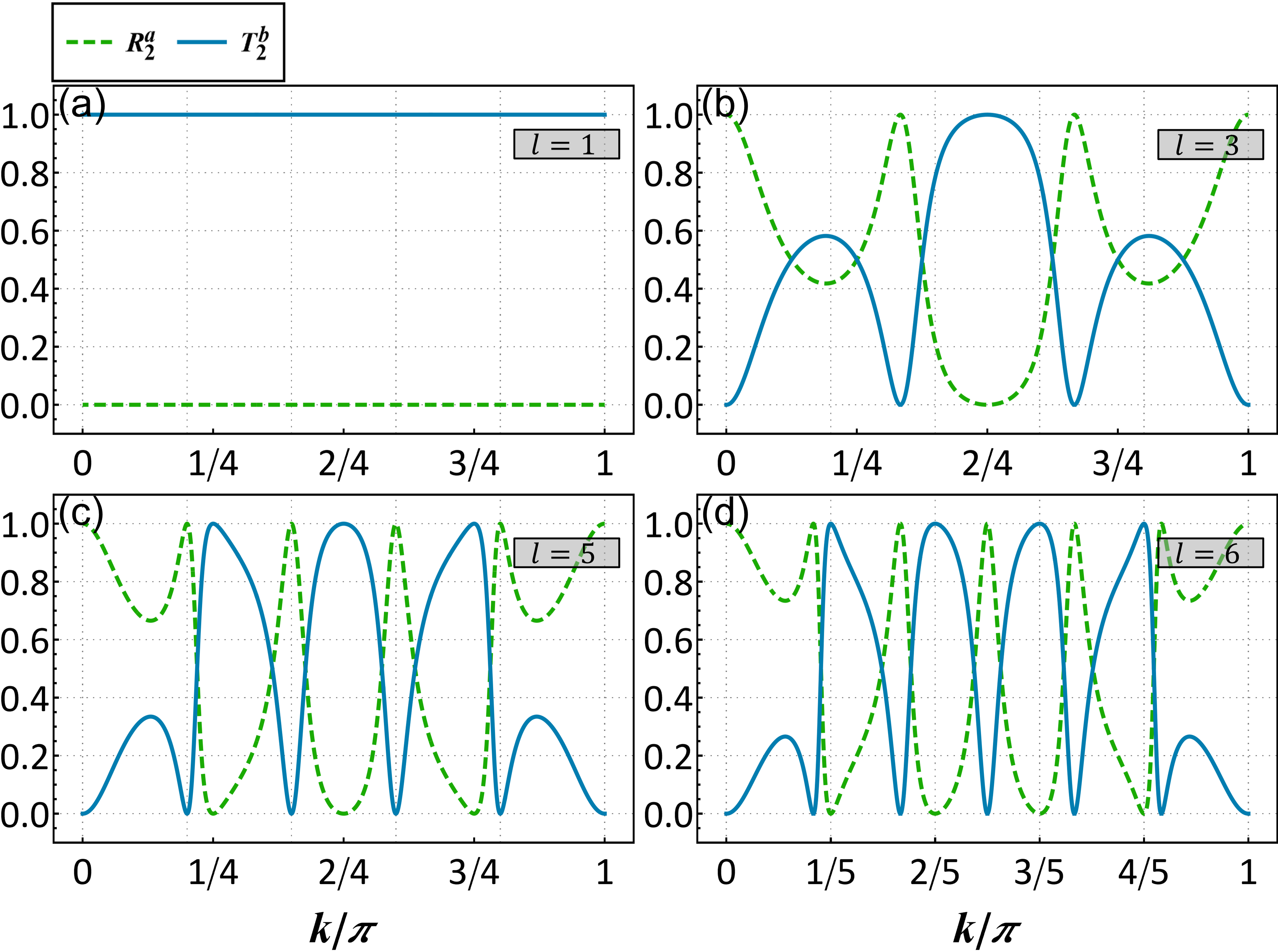}
	
	\caption{(color online) Reflection rate (green dashed line) $R_2^a$ and transfer rate (blue solid line) $T_2^b$ as functions of the wave number of the incident wave $k/\pi$ ($0\le k\le\pi$, the diagram within the $\pi \le k\le 2\pi$ is exactly the same), considering different coupling points of the giant atom with CRW-a at cavity $a_l$.
		(a) $l=1$.
		(b) $l=3$.
		(c) $l=5$.
		(d) $l=6$.
		The values of the other parameters remain consistent with the ones shown in Fig. \ref{fig2}(d), namely, $g_a=g_b=\Omega=1$.}
	
	\label{figs3}
\end{figure}

In Fig. \ref{figs3}, we present curves depicting the variations of reflection and transfer rates with the wave number $k/\pi$ for different cavity coupling strengths of the giant atom with CRW-a. When $l=1$, as mentioned earlier, it corresponds to the model depicted in Fig. \ref{fig1} and also yields identical results, referring to Fig. \ref{figs3}(a). From Figs. \ref{figs3}(b) to \ref{figs3}(d), we observe that when $k$ satisfies $k=\frac{n\pi}{l-1}\left(0\le k\le\pi\right)$, the reflection rate is 0, where $n=1,2,\cdots,n$ represents all positive integers.





To explain the observed phenomenon in Fig. \ref{figs3} using the pathways interference mechanism proposed above, let's first analyze the single-photon scattering process as depicted in Fig. \ref{figs2}. When a plane wave propagating freely to the right along CRW-a encounters the scattering cavity $a_l$, it undergoes the first scattering, resulting in reflection, transmission, and transfer waves. The reflection wave propagates freely to the left along CRW-a. The transfer wave propagates to the left along CRW-b. And the transmission wave continues propagating along CRW-a until reaching cavity $a_1$, where it is reflected due to the boundary, creating a boundary reflection wave. This boundary reflection wave propagates to the left along CRW-a and undergoes the second scattering when it encounters cavity $a_l$, resulting in reflection, transmission, and transfer waves again. Of note, the presence of hard wall boundary condition at the end of CRW-a causes cavity $a_1$ to behave like a perfect mirror. Thus, the second scattering discussed here is absent from the model depicted in Fig. \ref{fig1}, where the cavity $a_1$ cannot serve as both a scattering cavity and a perfect mirror.

During the first scattering process, the incident wave from CRW-a either skims directly over the giant atom, or is captured by the giant atom and re-emitted into three channels due to the influence of spontaneous emission of giant atom. Among these, the photon absorbed and re-emitted into channel I make up a part of the final reflection wave, as depicted in Fig. \ref{figs2}.
In the second scattering process, the boundary reflection wave within channel II either directly bypasses the giant atom or is captured by the giant atom and subsequently re-emitted into three channels when propagating to cavity $a_l$. Among these, the photon that skim directly over the giant atom and those absorbed and re-emitted into channel I make up the other part of the final reflection wave, with both of them having identical phases.
Hence, the final reflection wave is effectively composed of three different waves superimposed: (i) wave 1, generated as the incident wave is absorbed by the giant atom and re-emitted into Channel I; (ii) wave 2a, which propagates along the positive $j_a$ axis, bypassing the giant atom in the boundary reflection wave; and (iii) wave 2b, absorbed and re-emitted by the giant atom into Channel I within the boundary reflection wave. With the constant phase difference between wave 2a and wave 2b, we designate the resultant constructively interfering wave as wave 2.

\begin{figure}[htb]
	\centering
	\includegraphics[width=12cm]{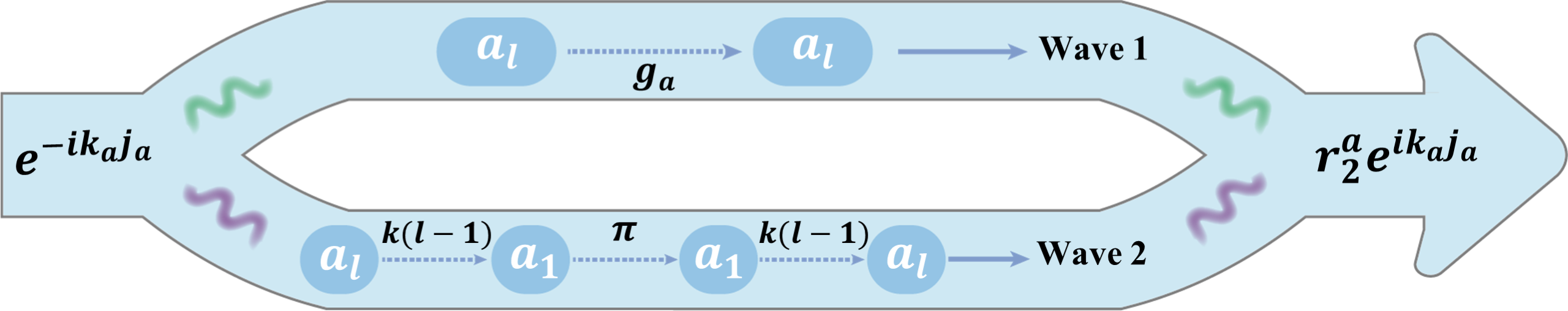}
	
	\caption{Schematic diagram illustrating the phase variations of the two waves composing the final reflected wave during the scattering process. Green waveform corresponds to wave 1, while the purple waveform represents wave 2.}
	
	\label{figs4}
\end{figure}

Next, we delve into the phase difference between wave 1 and wave 2 when they form the final reflected wave. Illustrated in Fig. \ref{figs4}, when the incident wave propagates to cavity $a_l$, it undergoes the first scattering, at which point we assume its phase to be $\varphi_0$. A portion of it, serving as wave 1, is absorbed by the giant atom and re-emitted into Channel I, maintaining a phase of $\varphi_0$. On the other hand, wave 2 functions as the transmission wave during the first scattering, continuing its rightward propagation along CRW-a. Upon reaching cavity $a_1$, it undergoes a phase shift of $k(l-1)$, followed by perfect reflection with an additional phase shift of $\pi$, and then it returns from cavity $a_1$ to $a_l$, incurring another phase shift of $k(l-1)$. Therefore, the phase of wave 2 is $\varphi_0+2k(l-1)+\pi$, and the phase difference between the two is:
\begin{equation}\label{eq14}
	\Delta \varphi=2k\left(l-1\right)+\pi,
\end{equation}
from Eq. \eqref{eq14}, it can be observed that when $l=1$, which means the boundary cavity of CRW-a is coupled to the giant atom, $\Delta\varphi\equiv\pi$ and is independent of the value of $k$. In this scenario, destructive interference occurs when waves 1 and 2 happen to have precisely the identical amplitudes, leading to the complete disappearance of the final reflected wave. Consequently, photon is all transferred to the other CRW, resulting in the perfect transfer, as shown in Fig. \ref{fig2}(d) and Fig. \ref{figs3}(a). The repeated reference to the parameter settings $g_a=g_b=\Omega=\xi$ above precisely ensures that these two reflected waves maintain identical amplitudes.
However, when $l$ takes other values, perfect transfer occurs only when $\Delta\varphi=\left(2n+1\right)\pi$, which corresponds to $k=\frac{n\pi}{l-1}$. This condition aligns with the observations in Fig. \ref{figs3}, thereby substantiating the physical mechanism explanation about pathway interference above.
This quantum interference phenomenon arising from the phase difference between different pathways is similar to the discussion in Section III of Ref.\cite{SQC_GA_1}, where the position-dependent phase resulting from the separation between cavity and emitter can also adjust the constructive and destructive interferences between two pathways, serving the purpose of modulating the nonreciprocal transport of single photon. 

\bibliography{refs}

\begin{thebibliography}{63}%
\makeatletter
\providecommand \@ifxundefined [1]{%
 \@ifx{#1\undefined}
}%
\providecommand \@ifnum [1]{%
 \ifnum #1\expandafter \@firstoftwo
 \else \expandafter \@secondoftwo
 \fi
}%
\providecommand \@ifx [1]{%
 \ifx #1\expandafter \@firstoftwo
 \else \expandafter \@secondoftwo
 \fi
}%
\providecommand \natexlab [1]{#1}%
\providecommand \enquote  [1]{``#1''}%
\providecommand \bibnamefont  [1]{#1}%
\providecommand \bibfnamefont [1]{#1}%
\providecommand \citenamefont [1]{#1}%
\providecommand \href@noop [0]{\@secondoftwo}%
\providecommand \href [0]{\begingroup \@sanitize@url \@href}%
\providecommand \@href[1]{\@@startlink{#1}\@@href}%
\providecommand \@@href[1]{\endgroup#1\@@endlink}%
\providecommand \@sanitize@url [0]{\catcode `\\12\catcode `\$12\catcode
  `\&12\catcode `\#12\catcode `\^12\catcode `\_12\catcode `\%12\relax}%
\providecommand \@@startlink[1]{}%
\providecommand \@@endlink[0]{}%
\providecommand \url  [0]{\begingroup\@sanitize@url \@url }%
\providecommand \@url [1]{\endgroup\@href {#1}{\urlprefix }}%
\providecommand \urlprefix  [0]{URL }%
\providecommand \Eprint [0]{\href }%
\providecommand \doibase [0]{https://doi.org/}%
\providecommand \selectlanguage [0]{\@gobble}%
\providecommand \bibinfo  [0]{\@secondoftwo}%
\providecommand \bibfield  [0]{\@secondoftwo}%
\providecommand \translation [1]{[#1]}%
\providecommand \BibitemOpen [0]{}%
\providecommand \bibitemStop [0]{}%
\providecommand \bibitemNoStop [0]{.\EOS\space}%
\providecommand \EOS [0]{\spacefactor3000\relax}%
\providecommand \BibitemShut  [1]{\csname bibitem#1\endcsname}%
\let\auto@bib@innerbib\@empty
\bibitem [{\citenamefont {Kimble}(2008)}]{quantum_network_1}%
  \BibitemOpen
  \bibfield  {author} {\bibinfo {author} {\bibfnamefont {H.~J.}\ \bibnamefont
  {Kimble}},\ }\bibfield  {title} {\bibinfo {title} {The quantum internet},\
  }\href {https://doi.org/10.1038/nature07127} {\bibfield  {journal} {\bibinfo
  {journal} {Nature}\ }\textbf {\bibinfo {volume} {453}},\ \bibinfo {pages}
  {1023} (\bibinfo {year} {2008})}\BibitemShut {NoStop}%
\bibitem [{\citenamefont {Wehner}\ \emph {et~al.}(2018)\citenamefont {Wehner},
  \citenamefont {Elkouss},\ and\ \citenamefont {Hanson}}]{quantum_network_2}%
  \BibitemOpen
  \bibfield  {author} {\bibinfo {author} {\bibfnamefont {S.}~\bibnamefont
  {Wehner}}, \bibinfo {author} {\bibfnamefont {D.}~\bibnamefont {Elkouss}},\
  and\ \bibinfo {author} {\bibfnamefont {R.}~\bibnamefont {Hanson}},\
  }\bibfield  {title} {\bibinfo {title} {Quantum internet: A vision for the
  road ahead},\ }\href {https://doi.org/10.1126/science.aam9288} {\bibfield
  {journal} {\bibinfo  {journal} {Science}\ }\textbf {\bibinfo {volume}
  {362}},\ \bibinfo {pages} {eaam9288} (\bibinfo {year} {2018})}\BibitemShut
  {NoStop}%
\bibitem [{\citenamefont {Chiribella}\ \emph {et~al.}(2009)\citenamefont
  {Chiribella}, \citenamefont {D'Ariano},\ and\ \citenamefont
  {Perinotti}}]{quantum_network_3}%
  \BibitemOpen
  \bibfield  {author} {\bibinfo {author} {\bibfnamefont {G.}~\bibnamefont
  {Chiribella}}, \bibinfo {author} {\bibfnamefont {G.~M.}\ \bibnamefont
  {D'Ariano}},\ and\ \bibinfo {author} {\bibfnamefont {P.}~\bibnamefont
  {Perinotti}},\ }\bibfield  {title} {\bibinfo {title} {Theoretical framework
  for quantum networks},\ }\href {https://doi.org/10.1103/PhysRevA.80.022339}
  {\bibfield  {journal} {\bibinfo  {journal} {Phys. Rev. A}\ }\textbf {\bibinfo
  {volume} {80}},\ \bibinfo {pages} {022339} (\bibinfo {year}
  {2009})}\BibitemShut {NoStop}%
\bibitem [{\citenamefont {Lo~Piparo}\ \emph {et~al.}(2020)\citenamefont
  {Lo~Piparo}, \citenamefont {Hanks}, \citenamefont {Nemoto},\ and\
  \citenamefont {Munro}}]{quantum_network_4}%
  \BibitemOpen
  \bibfield  {author} {\bibinfo {author} {\bibfnamefont {N.}~\bibnamefont
  {Lo~Piparo}}, \bibinfo {author} {\bibfnamefont {M.}~\bibnamefont {Hanks}},
  \bibinfo {author} {\bibfnamefont {K.}~\bibnamefont {Nemoto}},\ and\ \bibinfo
  {author} {\bibfnamefont {W.~J.}\ \bibnamefont {Munro}},\ }\bibfield  {title}
  {\bibinfo {title} {Aggregating quantum networks},\ }\href
  {https://doi.org/10.1103/PhysRevA.102.052613} {\bibfield  {journal} {\bibinfo
   {journal} {Phys. Rev. A}\ }\textbf {\bibinfo {volume} {102}},\ \bibinfo
  {pages} {052613} (\bibinfo {year} {2020})}\BibitemShut {NoStop}%
\bibitem [{\citenamefont {Chang}\ \emph
  {et~al.}(2018{\natexlab{a}})\citenamefont {Chang}, \citenamefont {Douglas},
  \citenamefont {Gonz\'alez-Tudela}, \citenamefont {Hung},\ and\ \citenamefont
  {Kimble}}]{quantum_network_5}%
  \BibitemOpen
  \bibfield  {author} {\bibinfo {author} {\bibfnamefont {D.~E.}\ \bibnamefont
  {Chang}}, \bibinfo {author} {\bibfnamefont {J.~S.}\ \bibnamefont {Douglas}},
  \bibinfo {author} {\bibfnamefont {A.}~\bibnamefont {Gonz\'alez-Tudela}},
  \bibinfo {author} {\bibfnamefont {C.-L.}\ \bibnamefont {Hung}},\ and\
  \bibinfo {author} {\bibfnamefont {H.~J.}\ \bibnamefont {Kimble}},\ }\bibfield
   {title} {\bibinfo {title} {Colloquium: Quantum matter built from nanoscopic
  lattices of atoms and photons},\ }\href
  {https://doi.org/10.1103/RevModPhys.90.031002} {\bibfield  {journal}
  {\bibinfo  {journal} {Rev. Mod. Phys.}\ }\textbf {\bibinfo {volume} {90}},\
  \bibinfo {pages} {031002} (\bibinfo {year} {2018}{\natexlab{a}})}\BibitemShut
  {NoStop}%
\bibitem [{\citenamefont {Zoller}\ \emph {et~al.}(2005)\citenamefont {Zoller},
  \citenamefont {Beth}, \citenamefont {Binosi}, \citenamefont {Blatt},
  \citenamefont {Briegel}, \citenamefont {Bruss}, \citenamefont {Calarco},
  \citenamefont {Cirac}, \citenamefont {Deutsch}, \citenamefont {Eisert},
  \citenamefont {Ekert}, \citenamefont {Fabre}, \citenamefont {Gisin},
  \citenamefont {Grangiere}, \citenamefont {Grassl}, \citenamefont {Haroche},
  \citenamefont {Imamoglu}, \citenamefont {Karlson}, \citenamefont {Kempe},
  \citenamefont {Kouwenhoven}, \citenamefont {Kr{\"o}ll}, \citenamefont
  {Leuchs}, \citenamefont {Lewenstein}, \citenamefont {Loss}, \citenamefont
  {L{\"u}tkenhaus}, \citenamefont {Massar}, \citenamefont {Mooij},
  \citenamefont {Plenio}, \citenamefont {Polzik}, \citenamefont {Popescu},
  \citenamefont {Rempe}, \citenamefont {Sergienko}, \citenamefont {Suter},
  \citenamefont {Twamley}, \citenamefont {Wendin}, \citenamefont {Werner},
  \citenamefont {Winter}, \citenamefont {Wrachtrup},\ and\ \citenamefont
  {Zeilinger}}]{quantum_communication_1}%
  \BibitemOpen
  \bibfield  {author} {\bibinfo {author} {\bibfnamefont {P.}~\bibnamefont
  {Zoller}}, \bibinfo {author} {\bibfnamefont {T.}~\bibnamefont {Beth}},
  \bibinfo {author} {\bibfnamefont {D.}~\bibnamefont {Binosi}}, \bibinfo
  {author} {\bibfnamefont {R.}~\bibnamefont {Blatt}}, \bibinfo {author}
  {\bibfnamefont {H.}~\bibnamefont {Briegel}}, \bibinfo {author} {\bibfnamefont
  {D.}~\bibnamefont {Bruss}}, \bibinfo {author} {\bibfnamefont
  {T.}~\bibnamefont {Calarco}}, \bibinfo {author} {\bibfnamefont {J.~I.}\
  \bibnamefont {Cirac}}, \bibinfo {author} {\bibfnamefont {D.}~\bibnamefont
  {Deutsch}}, \bibinfo {author} {\bibfnamefont {J.}~\bibnamefont {Eisert}},
  \bibinfo {author} {\bibfnamefont {A.}~\bibnamefont {Ekert}}, \bibinfo
  {author} {\bibfnamefont {C.}~\bibnamefont {Fabre}}, \bibinfo {author}
  {\bibfnamefont {N.}~\bibnamefont {Gisin}}, \bibinfo {author} {\bibfnamefont
  {P.}~\bibnamefont {Grangiere}}, \bibinfo {author} {\bibfnamefont
  {M.}~\bibnamefont {Grassl}}, \bibinfo {author} {\bibfnamefont
  {S.}~\bibnamefont {Haroche}}, \bibinfo {author} {\bibfnamefont
  {A.}~\bibnamefont {Imamoglu}}, \bibinfo {author} {\bibfnamefont
  {A.}~\bibnamefont {Karlson}}, \bibinfo {author} {\bibfnamefont
  {J.}~\bibnamefont {Kempe}}, \bibinfo {author} {\bibfnamefont
  {L.}~\bibnamefont {Kouwenhoven}}, \bibinfo {author} {\bibfnamefont
  {S.}~\bibnamefont {Kr{\"o}ll}}, \bibinfo {author} {\bibfnamefont
  {G.}~\bibnamefont {Leuchs}}, \bibinfo {author} {\bibfnamefont
  {M.}~\bibnamefont {Lewenstein}}, \bibinfo {author} {\bibfnamefont
  {D.}~\bibnamefont {Loss}}, \bibinfo {author} {\bibfnamefont {N.}~\bibnamefont
  {L{\"u}tkenhaus}}, \bibinfo {author} {\bibfnamefont {S.}~\bibnamefont
  {Massar}}, \bibinfo {author} {\bibfnamefont {J.~E.}\ \bibnamefont {Mooij}},
  \bibinfo {author} {\bibfnamefont {M.~B.}\ \bibnamefont {Plenio}}, \bibinfo
  {author} {\bibfnamefont {E.}~\bibnamefont {Polzik}}, \bibinfo {author}
  {\bibfnamefont {S.}~\bibnamefont {Popescu}}, \bibinfo {author} {\bibfnamefont
  {G.}~\bibnamefont {Rempe}}, \bibinfo {author} {\bibfnamefont
  {A.}~\bibnamefont {Sergienko}}, \bibinfo {author} {\bibfnamefont
  {D.}~\bibnamefont {Suter}}, \bibinfo {author} {\bibfnamefont
  {J.}~\bibnamefont {Twamley}}, \bibinfo {author} {\bibfnamefont
  {G.}~\bibnamefont {Wendin}}, \bibinfo {author} {\bibfnamefont
  {R.}~\bibnamefont {Werner}}, \bibinfo {author} {\bibfnamefont
  {A.}~\bibnamefont {Winter}}, \bibinfo {author} {\bibfnamefont
  {J.}~\bibnamefont {Wrachtrup}},\ and\ \bibinfo {author} {\bibfnamefont
  {A.}~\bibnamefont {Zeilinger}},\ }\bibfield  {title} {\bibinfo {title}
  {Quantum information processing and communication},\ }\href
  {https://doi.org/10.1140/epjd/e2005-00251-1} {\bibfield  {journal} {\bibinfo
  {journal} {Eur. Phys. J. D - Atomic, Molecular, Optical and Plasma Physics}\
  }\textbf {\bibinfo {volume} {36}},\ \bibinfo {pages} {203} (\bibinfo {year}
  {2005})}\BibitemShut {NoStop}%
\bibitem [{\citenamefont {Bennett}\ and\ \citenamefont
  {Brassard}(2014)}]{quantum_communication_2}%
  \BibitemOpen
  \bibfield  {author} {\bibinfo {author} {\bibfnamefont {C.~H.}\ \bibnamefont
  {Bennett}}\ and\ \bibinfo {author} {\bibfnamefont {G.}~\bibnamefont
  {Brassard}},\ }\bibfield  {title} {\bibinfo {title} {Quantum cryptography:
  Public key distribution and coin tossing},\ }\href
  {https://doi.org/10.1016/j.tcs.2014.05.025} {\bibfield  {journal} {\bibinfo
  {journal} {Theor Comput Sci}\ }\textbf {\bibinfo {volume} {560}},\ \bibinfo
  {pages} {7} (\bibinfo {year} {2014})},\ \bibinfo {note} {theoretical Aspects
  of Quantum Cryptography – celebrating 30 years of BB84}\BibitemShut
  {NoStop}%
\bibitem [{\citenamefont {Qiu}(2014)}]{quantum_communication_3}%
  \BibitemOpen
  \bibfield  {author} {\bibinfo {author} {\bibfnamefont {J.}~\bibnamefont
  {Qiu}},\ }\bibfield  {title} {\bibinfo {title} {Quantum communications leap
  out of the lab},\ }\href {https://doi.org/10.1038/508441a} {\bibfield
  {journal} {\bibinfo  {journal} {Nature}\ }\textbf {\bibinfo {volume} {508}},\
  \bibinfo {pages} {441} (\bibinfo {year} {2014})}\BibitemShut {NoStop}%
\bibitem [{\citenamefont {Bennett}\ and\ \citenamefont
  {DiVincenzo}(2000)}]{quantum_communication_4}%
  \BibitemOpen
  \bibfield  {author} {\bibinfo {author} {\bibfnamefont {C.~H.}\ \bibnamefont
  {Bennett}}\ and\ \bibinfo {author} {\bibfnamefont {D.~P.}\ \bibnamefont
  {DiVincenzo}},\ }\bibfield  {title} {\bibinfo {title} {Quantum information
  and computation},\ }\href {https://doi.org/10.1038/35005001} {\bibfield
  {journal} {\bibinfo  {journal} {Nature}\ }\textbf {\bibinfo {volume} {404}},\
  \bibinfo {pages} {247} (\bibinfo {year} {2000})}\BibitemShut {NoStop}%
\bibitem [{\citenamefont {Duan}\ and\ \citenamefont
  {Raussendorf}(2005)}]{quantum_computing_1}%
  \BibitemOpen
  \bibfield  {author} {\bibinfo {author} {\bibfnamefont {L.-M.}\ \bibnamefont
  {Duan}}\ and\ \bibinfo {author} {\bibfnamefont {R.}~\bibnamefont
  {Raussendorf}},\ }\bibfield  {title} {\bibinfo {title} {Efficient quantum
  computation with probabilistic quantum gates},\ }\href
  {https://doi.org/10.1103/PhysRevLett.95.080503} {\bibfield  {journal}
  {\bibinfo  {journal} {Phys. Rev. Lett.}\ }\textbf {\bibinfo {volume} {95}},\
  \bibinfo {pages} {080503} (\bibinfo {year} {2005})}\BibitemShut {NoStop}%
\bibitem [{\citenamefont {Giovannetti}\ \emph {et~al.}(2011)\citenamefont
  {Giovannetti}, \citenamefont {Lloyd},\ and\ \citenamefont
  {Maccone}}]{quantum_metrology_1}%
  \BibitemOpen
  \bibfield  {author} {\bibinfo {author} {\bibfnamefont {V.}~\bibnamefont
  {Giovannetti}}, \bibinfo {author} {\bibfnamefont {S.}~\bibnamefont {Lloyd}},\
  and\ \bibinfo {author} {\bibfnamefont {L.}~\bibnamefont {Maccone}},\
  }\bibfield  {title} {\bibinfo {title} {Advances in quantum metrology},\
  }\href {https://doi.org/10.1038/nphoton.2011.35} {\bibfield  {journal}
  {\bibinfo  {journal} {Nat. Photonics}\ }\textbf {\bibinfo {volume} {5}},\
  \bibinfo {pages} {222} (\bibinfo {year} {2011})}\BibitemShut {NoStop}%
\bibitem [{\citenamefont {Giovannetti}\ \emph {et~al.}(2006)\citenamefont
  {Giovannetti}, \citenamefont {Lloyd},\ and\ \citenamefont
  {Maccone}}]{quantum_metrology_2}%
  \BibitemOpen
  \bibfield  {author} {\bibinfo {author} {\bibfnamefont {V.}~\bibnamefont
  {Giovannetti}}, \bibinfo {author} {\bibfnamefont {S.}~\bibnamefont {Lloyd}},\
  and\ \bibinfo {author} {\bibfnamefont {L.}~\bibnamefont {Maccone}},\
  }\bibfield  {title} {\bibinfo {title} {Quantum metrology},\ }\href
  {https://doi.org/10.1103/PhysRevLett.96.010401} {\bibfield  {journal}
  {\bibinfo  {journal} {Phys. Rev. Lett.}\ }\textbf {\bibinfo {volume} {96}},\
  \bibinfo {pages} {010401} (\bibinfo {year} {2006})}\BibitemShut {NoStop}%
\bibitem [{\citenamefont {Zhou}\ \emph
  {et~al.}(2008{\natexlab{a}})\citenamefont {Zhou}, \citenamefont {Dong},
  \citenamefont {Liu}, \citenamefont {Sun},\ and\ \citenamefont
  {Nori}}]{information_carriers_router_1}%
  \BibitemOpen
  \bibfield  {author} {\bibinfo {author} {\bibfnamefont {L.}~\bibnamefont
  {Zhou}}, \bibinfo {author} {\bibfnamefont {H.}~\bibnamefont {Dong}}, \bibinfo
  {author} {\bibfnamefont {Y.-x.}\ \bibnamefont {Liu}}, \bibinfo {author}
  {\bibfnamefont {C.~P.}\ \bibnamefont {Sun}},\ and\ \bibinfo {author}
  {\bibfnamefont {F.}~\bibnamefont {Nori}},\ }\bibfield  {title} {\bibinfo
  {title} {Quantum supercavity with atomic mirrors},\ }\href
  {https://doi.org/10.1103/PhysRevA.78.063827} {\bibfield  {journal} {\bibinfo
  {journal} {Phys. Rev. A}\ }\textbf {\bibinfo {volume} {78}},\ \bibinfo
  {pages} {063827} (\bibinfo {year} {2008}{\natexlab{a}})}\BibitemShut
  {NoStop}%
\bibitem [{\citenamefont {Yan}\ and\ \citenamefont
  {Fan}(2014)}]{information_carriers_router_2}%
  \BibitemOpen
  \bibfield  {author} {\bibinfo {author} {\bibfnamefont {W.-B.}\ \bibnamefont
  {Yan}}\ and\ \bibinfo {author} {\bibfnamefont {H.}~\bibnamefont {Fan}},\
  }\bibfield  {title} {\bibinfo {title} {Control of single-photon transport in
  a one-dimensional waveguide by a single photon},\ }\href
  {https://doi.org/10.1103/PhysRevA.90.053807} {\bibfield  {journal} {\bibinfo
  {journal} {Phys. Rev. A}\ }\textbf {\bibinfo {volume} {90}},\ \bibinfo
  {pages} {053807} (\bibinfo {year} {2014})}\BibitemShut {NoStop}%
\bibitem [{\citenamefont {Zhou}\ \emph
  {et~al.}(2008{\natexlab{b}})\citenamefont {Zhou}, \citenamefont {Gong},
  \citenamefont {Liu}, \citenamefont {Sun},\ and\ \citenamefont
  {Nori}}]{1_d_crw_router}%
  \BibitemOpen
  \bibfield  {author} {\bibinfo {author} {\bibfnamefont {L.}~\bibnamefont
  {Zhou}}, \bibinfo {author} {\bibfnamefont {Z.~R.}\ \bibnamefont {Gong}},
  \bibinfo {author} {\bibfnamefont {Y.-x.}\ \bibnamefont {Liu}}, \bibinfo
  {author} {\bibfnamefont {C.~P.}\ \bibnamefont {Sun}},\ and\ \bibinfo {author}
  {\bibfnamefont {F.}~\bibnamefont {Nori}},\ }\bibfield  {title} {\bibinfo
  {title} {Controllable scattering of a single photon inside a one-dimensional
  resonator waveguide},\ }\href
  {https://doi.org/10.1103/PhysRevLett.101.100501} {\bibfield  {journal}
  {\bibinfo  {journal} {Phys. Rev. Lett.}\ }\textbf {\bibinfo {volume} {101}},\
  \bibinfo {pages} {100501} (\bibinfo {year} {2008}{\natexlab{b}})}\BibitemShut
  {NoStop}%
\bibitem [{\citenamefont {Wu}\ \emph {et~al.}(2022)\citenamefont {Wu},
  \citenamefont {Dong}, \citenamefont {Xu}, \citenamefont {Zou},\ and\
  \citenamefont {Zhang}}]{optical_router}%
  \BibitemOpen
  \bibfield  {author} {\bibinfo {author} {\bibfnamefont {J.-N.}\ \bibnamefont
  {Wu}}, \bibinfo {author} {\bibfnamefont {J.}~\bibnamefont {Dong}}, \bibinfo
  {author} {\bibfnamefont {Y.}~\bibnamefont {Xu}}, \bibinfo {author}
  {\bibfnamefont {B.}~\bibnamefont {Zou}},\ and\ \bibinfo {author}
  {\bibfnamefont {Y.}~\bibnamefont {Zhang}},\ }\bibfield  {title} {\bibinfo
  {title} {Multichannel adjustable single-photon router based on large
  detuning},\ }\href {https://doi.org/10.1103/PhysRevApplied.18.054007}
  {\bibfield  {journal} {\bibinfo  {journal} {Phys. Rev. Appl.}\ }\textbf
  {\bibinfo {volume} {18}},\ \bibinfo {pages} {054007} (\bibinfo {year}
  {2022})}\BibitemShut {NoStop}%
\bibitem [{\citenamefont {Zhou}\ \emph {et~al.}(2009)\citenamefont {Zhou},
  \citenamefont {Yang}, \citenamefont {Liu}, \citenamefont {Sun},\ and\
  \citenamefont {Nori}}]{quantum_zeno_switch}%
  \BibitemOpen
  \bibfield  {author} {\bibinfo {author} {\bibfnamefont {L.}~\bibnamefont
  {Zhou}}, \bibinfo {author} {\bibfnamefont {S.}~\bibnamefont {Yang}}, \bibinfo
  {author} {\bibfnamefont {Y.-x.}\ \bibnamefont {Liu}}, \bibinfo {author}
  {\bibfnamefont {C.~P.}\ \bibnamefont {Sun}},\ and\ \bibinfo {author}
  {\bibfnamefont {F.}~\bibnamefont {Nori}},\ }\bibfield  {title} {\bibinfo
  {title} {Quantum zeno switch for single-photon coherent transport},\ }\href
  {https://doi.org/10.1103/PhysRevA.80.062109} {\bibfield  {journal} {\bibinfo
  {journal} {Phys. Rev. A}\ }\textbf {\bibinfo {volume} {80}},\ \bibinfo
  {pages} {062109} (\bibinfo {year} {2009})}\BibitemShut {NoStop}%
\bibitem [{\citenamefont {Gonzalez-Ballestero}\ \emph
  {et~al.}(2016)\citenamefont {Gonzalez-Ballestero}, \citenamefont {Moreno},
  \citenamefont {Garcia-Vidal},\ and\ \citenamefont
  {Gonzalez-Tudela}}]{two_waveguide_atom_2}%
  \BibitemOpen
  \bibfield  {author} {\bibinfo {author} {\bibfnamefont {C.}~\bibnamefont
  {Gonzalez-Ballestero}}, \bibinfo {author} {\bibfnamefont {E.}~\bibnamefont
  {Moreno}}, \bibinfo {author} {\bibfnamefont {F.~J.}\ \bibnamefont
  {Garcia-Vidal}},\ and\ \bibinfo {author} {\bibfnamefont {A.}~\bibnamefont
  {Gonzalez-Tudela}},\ }\bibfield  {title} {\bibinfo {title} {Nonreciprocal
  few-photon routing schemes based on chiral waveguide-emitter couplings},\
  }\href {https://doi.org/10.1103/PhysRevA.94.063817} {\bibfield  {journal}
  {\bibinfo  {journal} {Phys. Rev. A}\ }\textbf {\bibinfo {volume} {94}},\
  \bibinfo {pages} {063817} (\bibinfo {year} {2016})}\BibitemShut {NoStop}%
\bibitem [{\citenamefont {Cheng}\ \emph {et~al.}(2016)\citenamefont {Cheng},
  \citenamefont {Ma}, \citenamefont {Zhang},\ and\ \citenamefont
  {Wang}}]{two_waveguide_atom_3}%
  \BibitemOpen
  \bibfield  {author} {\bibinfo {author} {\bibfnamefont {M.-T.}\ \bibnamefont
  {Cheng}}, \bibinfo {author} {\bibfnamefont {X.-S.}\ \bibnamefont {Ma}},
  \bibinfo {author} {\bibfnamefont {J.-Y.}\ \bibnamefont {Zhang}},\ and\
  \bibinfo {author} {\bibfnamefont {B.}~\bibnamefont {Wang}},\ }\bibfield
  {title} {\bibinfo {title} {Single photon transport in two waveguides chirally
  coupled by a quantum emitter},\ }\href {https://doi.org/10.1364/OE.24.019988}
  {\bibfield  {journal} {\bibinfo  {journal} {Opt. Express}\ }\textbf {\bibinfo
  {volume} {24}},\ \bibinfo {pages} {19988} (\bibinfo {year}
  {2016})}\BibitemShut {NoStop}%
\bibitem [{\citenamefont {Zhou}\ \emph {et~al.}(2013)\citenamefont {Zhou},
  \citenamefont {Yang}, \citenamefont {Li},\ and\ \citenamefont
  {Sun}}]{x_shape_router_1}%
  \BibitemOpen
  \bibfield  {author} {\bibinfo {author} {\bibfnamefont {L.}~\bibnamefont
  {Zhou}}, \bibinfo {author} {\bibfnamefont {L.-P.}\ \bibnamefont {Yang}},
  \bibinfo {author} {\bibfnamefont {Y.}~\bibnamefont {Li}},\ and\ \bibinfo
  {author} {\bibfnamefont {C.~P.}\ \bibnamefont {Sun}},\ }\bibfield  {title}
  {\bibinfo {title} {Quantum routing of single photons with a cyclic
  three-level system},\ }\href {https://doi.org/10.1103/PhysRevLett.111.103604}
  {\bibfield  {journal} {\bibinfo  {journal} {Phys. Rev. Lett.}\ }\textbf
  {\bibinfo {volume} {111}},\ \bibinfo {pages} {103604} (\bibinfo {year}
  {2013})}\BibitemShut {NoStop}%
\bibitem [{\citenamefont {Lu}\ \emph {et~al.}(2014)\citenamefont {Lu},
  \citenamefont {Zhou}, \citenamefont {Kuang},\ and\ \citenamefont
  {Nori}}]{x_shape_router_2}%
  \BibitemOpen
  \bibfield  {author} {\bibinfo {author} {\bibfnamefont {J.}~\bibnamefont
  {Lu}}, \bibinfo {author} {\bibfnamefont {L.}~\bibnamefont {Zhou}}, \bibinfo
  {author} {\bibfnamefont {L.-M.}\ \bibnamefont {Kuang}},\ and\ \bibinfo
  {author} {\bibfnamefont {F.}~\bibnamefont {Nori}},\ }\bibfield  {title}
  {\bibinfo {title} {Single-photon router: Coherent control of multichannel
  scattering for single photons with quantum interferences},\ }\href
  {https://doi.org/10.1103/PhysRevA.89.013805} {\bibfield  {journal} {\bibinfo
  {journal} {Phys. Rev. A}\ }\textbf {\bibinfo {volume} {89}},\ \bibinfo
  {pages} {013805} (\bibinfo {year} {2014})}\BibitemShut {NoStop}%
\bibitem [{\citenamefont {Lu}\ \emph {et~al.}(2015)\citenamefont {Lu},
  \citenamefont {Wang},\ and\ \citenamefont {Zhou}}]{t_shape_router_1}%
  \BibitemOpen
  \bibfield  {author} {\bibinfo {author} {\bibfnamefont {J.}~\bibnamefont
  {Lu}}, \bibinfo {author} {\bibfnamefont {Z.~H.}\ \bibnamefont {Wang}},\ and\
  \bibinfo {author} {\bibfnamefont {L.}~\bibnamefont {Zhou}},\ }\bibfield
  {title} {\bibinfo {title} {T-shaped single-photon router},\ }\href
  {https://doi.org/10.1364/OE.23.022955} {\bibfield  {journal} {\bibinfo
  {journal} {Opt. Express}\ }\textbf {\bibinfo {volume} {23}},\ \bibinfo
  {pages} {22955} (\bibinfo {year} {2015})}\BibitemShut {NoStop}%
\bibitem [{\citenamefont {Korzeczek}\ and\ \citenamefont
  {Braun}(2021)}]{atomic_ensemble_router}%
  \BibitemOpen
  \bibfield  {author} {\bibinfo {author} {\bibfnamefont {M.~C.}\ \bibnamefont
  {Korzeczek}}\ and\ \bibinfo {author} {\bibfnamefont {D.}~\bibnamefont
  {Braun}},\ }\bibfield  {title} {\bibinfo {title} {Quantum router: Storing and
  redirecting light at the photon level},\ }\href
  {https://doi.org/10.1103/PhysRevA.104.063714} {\bibfield  {journal} {\bibinfo
   {journal} {Phys. Rev. A}\ }\textbf {\bibinfo {volume} {104}},\ \bibinfo
  {pages} {063714} (\bibinfo {year} {2021})}\BibitemShut {NoStop}%
\bibitem [{\citenamefont {Ren}\ \emph {et~al.}(2022)\citenamefont {Ren},
  \citenamefont {Ma}, \citenamefont {Xie}, \citenamefont {Li}, \citenamefont
  {Cao},\ and\ \citenamefont {Li}}]{superconducting_circuit_router_1}%
  \BibitemOpen
  \bibfield  {author} {\bibinfo {author} {\bibfnamefont {Y.-l.}\ \bibnamefont
  {Ren}}, \bibinfo {author} {\bibfnamefont {S.-l.}\ \bibnamefont {Ma}},
  \bibinfo {author} {\bibfnamefont {J.-k.}\ \bibnamefont {Xie}}, \bibinfo
  {author} {\bibfnamefont {X.-k.}\ \bibnamefont {Li}}, \bibinfo {author}
  {\bibfnamefont {M.-t.}\ \bibnamefont {Cao}},\ and\ \bibinfo {author}
  {\bibfnamefont {F.-l.}\ \bibnamefont {Li}},\ }\bibfield  {title} {\bibinfo
  {title} {Nonreciprocal single-photon quantum router},\ }\href
  {https://doi.org/10.1103/PhysRevA.105.013711} {\bibfield  {journal} {\bibinfo
   {journal} {Phys. Rev. A}\ }\textbf {\bibinfo {volume} {105}},\ \bibinfo
  {pages} {013711} (\bibinfo {year} {2022})}\BibitemShut {NoStop}%
\bibitem [{\citenamefont {Wang}\ \emph
  {et~al.}(2021{\natexlab{a}})\citenamefont {Wang}, \citenamefont {Wu},
  \citenamefont {Bao}, \citenamefont {Li}, \citenamefont {Ma}, \citenamefont
  {Wang}, \citenamefont {Song}, \citenamefont {Zhang},\ and\ \citenamefont
  {Duan}}]{superconducting_circuit_router_2}%
  \BibitemOpen
  \bibfield  {author} {\bibinfo {author} {\bibfnamefont {Z.}~\bibnamefont
  {Wang}}, \bibinfo {author} {\bibfnamefont {Y.}~\bibnamefont {Wu}}, \bibinfo
  {author} {\bibfnamefont {Z.}~\bibnamefont {Bao}}, \bibinfo {author}
  {\bibfnamefont {Y.}~\bibnamefont {Li}}, \bibinfo {author} {\bibfnamefont
  {C.}~\bibnamefont {Ma}}, \bibinfo {author} {\bibfnamefont {H.}~\bibnamefont
  {Wang}}, \bibinfo {author} {\bibfnamefont {Y.}~\bibnamefont {Song}}, \bibinfo
  {author} {\bibfnamefont {H.}~\bibnamefont {Zhang}},\ and\ \bibinfo {author}
  {\bibfnamefont {L.}~\bibnamefont {Duan}},\ }\bibfield  {title} {\bibinfo
  {title} {Experimental realization of a deterministic quantum router with
  superconducting quantum circuits},\ }\href
  {https://doi.org/10.1103/PhysRevApplied.15.014049} {\bibfield  {journal}
  {\bibinfo  {journal} {Phys. Rev. Appl.}\ }\textbf {\bibinfo {volume} {15}},\
  \bibinfo {pages} {014049} (\bibinfo {year} {2021}{\natexlab{a}})}\BibitemShut
  {NoStop}%
\bibitem [{\citenamefont {Zhu}\ and\ \citenamefont
  {Jia}(2019)}]{superconducting_circuit_router_3}%
  \BibitemOpen
  \bibfield  {author} {\bibinfo {author} {\bibfnamefont {Y.~T.}\ \bibnamefont
  {Zhu}}\ and\ \bibinfo {author} {\bibfnamefont {W.~Z.}\ \bibnamefont {Jia}},\
  }\bibfield  {title} {\bibinfo {title} {Single-photon quantum router in the
  microwave regime utilizing double superconducting resonators with tunable
  coupling},\ }\href {https://doi.org/10.1103/PhysRevA.99.063815} {\bibfield
  {journal} {\bibinfo  {journal} {Phys. Rev. A}\ }\textbf {\bibinfo {volume}
  {99}},\ \bibinfo {pages} {063815} (\bibinfo {year} {2019})}\BibitemShut
  {NoStop}%
\bibitem [{\citenamefont {Agarwal}\ and\ \citenamefont
  {Huang}(2012)}]{optomechanical_system_router_1}%
  \BibitemOpen
  \bibfield  {author} {\bibinfo {author} {\bibfnamefont {G.~S.}\ \bibnamefont
  {Agarwal}}\ and\ \bibinfo {author} {\bibfnamefont {S.}~\bibnamefont
  {Huang}},\ }\bibfield  {title} {\bibinfo {title} {Optomechanical systems as
  single-photon routers},\ }\href {https://doi.org/10.1103/PhysRevA.85.021801}
  {\bibfield  {journal} {\bibinfo  {journal} {Phys. Rev. A}\ }\textbf {\bibinfo
  {volume} {85}},\ \bibinfo {pages} {021801} (\bibinfo {year}
  {2012})}\BibitemShut {NoStop}%
\bibitem [{\citenamefont {Du}\ \emph {et~al.}(2020)\citenamefont {Du},
  \citenamefont {Chen}, \citenamefont {Wu},\ and\ \citenamefont
  {Li}}]{optomechanical_system_router_2}%
  \BibitemOpen
  \bibfield  {author} {\bibinfo {author} {\bibfnamefont {L.}~\bibnamefont
  {Du}}, \bibinfo {author} {\bibfnamefont {Y.-T.}\ \bibnamefont {Chen}},
  \bibinfo {author} {\bibfnamefont {J.-H.}\ \bibnamefont {Wu}},\ and\ \bibinfo
  {author} {\bibfnamefont {Y.}~\bibnamefont {Li}},\ }\bibfield  {title}
  {\bibinfo {title} {Nonreciprocal interference and coherent photon routing in
  a three-port optomechanical system},\ }\href
  {https://doi.org/10.1364/OE.379990} {\bibfield  {journal} {\bibinfo
  {journal} {Opt. Express}\ }\textbf {\bibinfo {volume} {28}},\ \bibinfo
  {pages} {3647} (\bibinfo {year} {2020})}\BibitemShut {NoStop}%
\bibitem [{\citenamefont {Yang}\ \emph {et~al.}(2021)\citenamefont {Yang},
  \citenamefont {Qin}, \citenamefont {Zhang}, \citenamefont {Mao},
  \citenamefont {Wang},\ and\ \citenamefont
  {Long}}]{optomechanical_system_router_3}%
  \BibitemOpen
  \bibfield  {author} {\bibinfo {author} {\bibfnamefont {H.}~\bibnamefont
  {Yang}}, \bibinfo {author} {\bibfnamefont {G.}~\bibnamefont {Qin}}, \bibinfo
  {author} {\bibfnamefont {H.}~\bibnamefont {Zhang}}, \bibinfo {author}
  {\bibfnamefont {X.}~\bibnamefont {Mao}}, \bibinfo {author} {\bibfnamefont
  {M.}~\bibnamefont {Wang}},\ and\ \bibinfo {author} {\bibfnamefont
  {G.}~\bibnamefont {Long}},\ }\bibfield  {title} {\bibinfo {title} {Multimode
  interference induced optical routing in an optical microcavity},\ }\href
  {https://doi.org/10.1002/andp.202000506} {\bibfield  {journal} {\bibinfo
  {journal} {Ann Phys}\ }\textbf {\bibinfo {volume} {533}},\ \bibinfo {pages}
  {2000506} (\bibinfo {year} {2021})}\BibitemShut {NoStop}%
\bibitem [{\citenamefont {Xu}\ \emph {et~al.}(2017{\natexlab{a}})\citenamefont
  {Xu}, \citenamefont {Chen}, \citenamefont {Li},\ and\ \citenamefont
  {Liu}}]{two_semi_infinite_crw_router}%
  \BibitemOpen
  \bibfield  {author} {\bibinfo {author} {\bibfnamefont {X.-W.}\ \bibnamefont
  {Xu}}, \bibinfo {author} {\bibfnamefont {A.-X.}\ \bibnamefont {Chen}},
  \bibinfo {author} {\bibfnamefont {Y.}~\bibnamefont {Li}},\ and\ \bibinfo
  {author} {\bibfnamefont {Y.-x.}\ \bibnamefont {Liu}},\ }\bibfield  {title}
  {\bibinfo {title} {Single-photon nonreciprocal transport in one-dimensional
  coupled-resonator waveguides},\ }\href
  {https://doi.org/10.1103/PhysRevA.95.063808} {\bibfield  {journal} {\bibinfo
  {journal} {Phys. Rev. A}\ }\textbf {\bibinfo {volume} {95}},\ \bibinfo
  {pages} {063808} (\bibinfo {year} {2017}{\natexlab{a}})}\BibitemShut
  {NoStop}%
\bibitem [{\citenamefont {Ahumada}\ \emph {et~al.}(2019)\citenamefont
  {Ahumada}, \citenamefont {Orellana}, \citenamefont {Dom\'{\i}nguez-Adame},\
  and\ \citenamefont {Malyshev}}]{multi_coupled_points_crw_router}%
  \BibitemOpen
  \bibfield  {author} {\bibinfo {author} {\bibfnamefont {M.}~\bibnamefont
  {Ahumada}}, \bibinfo {author} {\bibfnamefont {P.~A.}\ \bibnamefont
  {Orellana}}, \bibinfo {author} {\bibfnamefont {F.}~\bibnamefont
  {Dom\'{\i}nguez-Adame}},\ and\ \bibinfo {author} {\bibfnamefont {A.~V.}\
  \bibnamefont {Malyshev}},\ }\bibfield  {title} {\bibinfo {title} {Tunable
  single-photon quantum router},\ }\href
  {https://doi.org/10.1103/PhysRevA.99.033827} {\bibfield  {journal} {\bibinfo
  {journal} {Phys. Rev. A}\ }\textbf {\bibinfo {volume} {99}},\ \bibinfo
  {pages} {033827} (\bibinfo {year} {2019})}\BibitemShut {NoStop}%
\bibitem [{\citenamefont {Plenio}\ and\ \citenamefont
  {Knight}(1998)}]{non_ideal_devices_1}%
  \BibitemOpen
  \bibfield  {author} {\bibinfo {author} {\bibfnamefont {M.~B.}\ \bibnamefont
  {Plenio}}\ and\ \bibinfo {author} {\bibfnamefont {P.~L.}\ \bibnamefont
  {Knight}},\ }\bibfield  {title} {\bibinfo {title} {The quantum-jump approach
  to dissipative dynamics in quantum optics},\ }\href
  {https://doi.org/10.1103/RevModPhys.70.101} {\bibfield  {journal} {\bibinfo
  {journal} {Rev. Mod. Phys.}\ }\textbf {\bibinfo {volume} {70}},\ \bibinfo
  {pages} {101} (\bibinfo {year} {1998})}\BibitemShut {NoStop}%
\bibitem [{\citenamefont {Chertkov}\ \emph {et~al.}(2023)\citenamefont
  {Chertkov}, \citenamefont {Cheng}, \citenamefont {Potter}, \citenamefont
  {Gopalakrishnan}, \citenamefont {Gatterman}, \citenamefont {Gerber},
  \citenamefont {Gilmore}, \citenamefont {Gresh}, \citenamefont {Hall},
  \citenamefont {Hankin}, \citenamefont {Matheny}, \citenamefont {Mengle},
  \citenamefont {Hayes}, \citenamefont {Neyenhuis}, \citenamefont {Stutz},\
  and\ \citenamefont {Foss-Feig}}]{non_ideal_devices_2}%
  \BibitemOpen
  \bibfield  {author} {\bibinfo {author} {\bibfnamefont {E.}~\bibnamefont
  {Chertkov}}, \bibinfo {author} {\bibfnamefont {Z.}~\bibnamefont {Cheng}},
  \bibinfo {author} {\bibfnamefont {A.~C.}\ \bibnamefont {Potter}}, \bibinfo
  {author} {\bibfnamefont {S.}~\bibnamefont {Gopalakrishnan}}, \bibinfo
  {author} {\bibfnamefont {T.~M.}\ \bibnamefont {Gatterman}}, \bibinfo {author}
  {\bibfnamefont {J.~A.}\ \bibnamefont {Gerber}}, \bibinfo {author}
  {\bibfnamefont {K.}~\bibnamefont {Gilmore}}, \bibinfo {author} {\bibfnamefont
  {D.}~\bibnamefont {Gresh}}, \bibinfo {author} {\bibfnamefont
  {A.}~\bibnamefont {Hall}}, \bibinfo {author} {\bibfnamefont {A.}~\bibnamefont
  {Hankin}}, \bibinfo {author} {\bibfnamefont {M.}~\bibnamefont {Matheny}},
  \bibinfo {author} {\bibfnamefont {T.}~\bibnamefont {Mengle}}, \bibinfo
  {author} {\bibfnamefont {D.}~\bibnamefont {Hayes}}, \bibinfo {author}
  {\bibfnamefont {B.}~\bibnamefont {Neyenhuis}}, \bibinfo {author}
  {\bibfnamefont {R.}~\bibnamefont {Stutz}},\ and\ \bibinfo {author}
  {\bibfnamefont {M.}~\bibnamefont {Foss-Feig}},\ }\bibfield  {title} {\bibinfo
  {title} {Characterizing a non-equilibrium phase transition on a quantum
  computer},\ }\bibfield  {journal} {\bibinfo  {journal} {Nat. Phys.}\ }\href
  {https://doi.org/10.1038/s41567-023-02199-w} {10.1038/s41567-023-02199-w}
  (\bibinfo {year} {2023})\BibitemShut {NoStop}%
\bibitem [{\citenamefont {Damanet}\ \emph {et~al.}(2019)\citenamefont
  {Damanet}, \citenamefont {Mascarenhas}, \citenamefont {Pekker},\ and\
  \citenamefont {Daley}}]{non_ideal_devices_3}%
  \BibitemOpen
  \bibfield  {author} {\bibinfo {author} {\bibfnamefont {F.~m.~c.}\
  \bibnamefont {Damanet}}, \bibinfo {author} {\bibfnamefont {E.}~\bibnamefont
  {Mascarenhas}}, \bibinfo {author} {\bibfnamefont {D.}~\bibnamefont
  {Pekker}},\ and\ \bibinfo {author} {\bibfnamefont {A.~J.}\ \bibnamefont
  {Daley}},\ }\bibfield  {title} {\bibinfo {title} {Controlling quantum
  transport via dissipation engineering},\ }\href
  {https://doi.org/10.1103/PhysRevLett.123.180402} {\bibfield  {journal}
  {\bibinfo  {journal} {Phys. Rev. Lett.}\ }\textbf {\bibinfo {volume} {123}},\
  \bibinfo {pages} {180402} (\bibinfo {year} {2019})}\BibitemShut {NoStop}%
\bibitem [{\citenamefont {Chang}\ \emph
  {et~al.}(2018{\natexlab{b}})\citenamefont {Chang}, \citenamefont {Douglas},
  \citenamefont {Gonz\'alez-Tudela}, \citenamefont {Hung},\ and\ \citenamefont
  {Kimble}}]{non_ideal_devices_4}%
  \BibitemOpen
  \bibfield  {author} {\bibinfo {author} {\bibfnamefont {D.~E.}\ \bibnamefont
  {Chang}}, \bibinfo {author} {\bibfnamefont {J.~S.}\ \bibnamefont {Douglas}},
  \bibinfo {author} {\bibfnamefont {A.}~\bibnamefont {Gonz\'alez-Tudela}},
  \bibinfo {author} {\bibfnamefont {C.-L.}\ \bibnamefont {Hung}},\ and\
  \bibinfo {author} {\bibfnamefont {H.~J.}\ \bibnamefont {Kimble}},\ }\bibfield
   {title} {\bibinfo {title} {Colloquium: Quantum matter built from nanoscopic
  lattices of atoms and photons},\ }\href
  {https://doi.org/10.1103/RevModPhys.90.031002} {\bibfield  {journal}
  {\bibinfo  {journal} {Rev. Mod. Phys.}\ }\textbf {\bibinfo {volume} {90}},\
  \bibinfo {pages} {031002} (\bibinfo {year} {2018}{\natexlab{b}})}\BibitemShut
  {NoStop}%
\bibitem [{\citenamefont {Makris}\ and\ \citenamefont
  {Christodoulides}(2006)}]{semi_infinite_crw}%
  \BibitemOpen
  \bibfield  {author} {\bibinfo {author} {\bibfnamefont {K.~G.}\ \bibnamefont
  {Makris}}\ and\ \bibinfo {author} {\bibfnamefont {D.~N.}\ \bibnamefont
  {Christodoulides}},\ }\bibfield  {title} {\bibinfo {title} {Method of images
  in optical discrete systems},\ }\href
  {https://doi.org/10.1103/PhysRevE.73.036616} {\bibfield  {journal} {\bibinfo
  {journal} {Phys. Rev. E}\ }\textbf {\bibinfo {volume} {73}},\ \bibinfo
  {pages} {036616} (\bibinfo {year} {2006})}\BibitemShut {NoStop}%
\bibitem [{\citenamefont {Xu}\ \emph {et~al.}(2017{\natexlab{b}})\citenamefont
  {Xu}, \citenamefont {Chen}, \citenamefont {Li},\ and\ \citenamefont
  {Liu}}]{multiple_semi_infinite_crw_router}%
  \BibitemOpen
  \bibfield  {author} {\bibinfo {author} {\bibfnamefont {X.-W.}\ \bibnamefont
  {Xu}}, \bibinfo {author} {\bibfnamefont {A.-X.}\ \bibnamefont {Chen}},
  \bibinfo {author} {\bibfnamefont {Y.}~\bibnamefont {Li}},\ and\ \bibinfo
  {author} {\bibfnamefont {Y.-x.}\ \bibnamefont {Liu}},\ }\bibfield  {title}
  {\bibinfo {title} {Nonreciprocal single-photon frequency converter via
  multiple semi-infinite coupled-resonator waveguides},\ }\href
  {https://doi.org/10.1103/PhysRevA.96.053853} {\bibfield  {journal} {\bibinfo
  {journal} {Phys. Rev. A}\ }\textbf {\bibinfo {volume} {96}},\ \bibinfo
  {pages} {053853} (\bibinfo {year} {2017}{\natexlab{b}})}\BibitemShut
  {NoStop}%
\bibitem [{\citenamefont {Gu}\ \emph {et~al.}(2017)\citenamefont {Gu},
  \citenamefont {Kockum}, \citenamefont {Miranowicz}, \citenamefont {xi~Liu},\
  and\ \citenamefont {Nori}}]{SQC_1}%
  \BibitemOpen
  \bibfield  {author} {\bibinfo {author} {\bibfnamefont {X.}~\bibnamefont
  {Gu}}, \bibinfo {author} {\bibfnamefont {A.~F.}\ \bibnamefont {Kockum}},
  \bibinfo {author} {\bibfnamefont {A.}~\bibnamefont {Miranowicz}}, \bibinfo
  {author} {\bibfnamefont {Y.}~\bibnamefont {xi~Liu}},\ and\ \bibinfo {author}
  {\bibfnamefont {F.}~\bibnamefont {Nori}},\ }\bibfield  {title} {\bibinfo
  {title} {Microwave photonics with superconducting quantum circuits},\ }\href
  {https://doi.org/https://doi.org/10.1016/j.physrep.2017.10.002} {\bibfield
  {journal} {\bibinfo  {journal} {Phys. Reports}\ }\textbf {\bibinfo {volume}
  {718-719}},\ \bibinfo {pages} {1} (\bibinfo {year} {2017})},\ \bibinfo {note}
  {microwave photonics with superconducting quantum circuits}\BibitemShut
  {NoStop}%
\bibitem [{\citenamefont {Blais}\ \emph {et~al.}(2020)\citenamefont {Blais},
  \citenamefont {Girvin},\ and\ \citenamefont {Oliver}}]{SQC_2}%
  \BibitemOpen
  \bibfield  {author} {\bibinfo {author} {\bibfnamefont {A.}~\bibnamefont
  {Blais}}, \bibinfo {author} {\bibfnamefont {S.~M.}\ \bibnamefont {Girvin}},\
  and\ \bibinfo {author} {\bibfnamefont {W.~D.}\ \bibnamefont {Oliver}},\
  }\bibfield  {title} {\bibinfo {title} {Quantum information processing and
  quantum optics with circuit quantum electrodynamics},\ }\href
  {https://doi.org/10.1038/s41567-020-0806-z} {\bibfield  {journal} {\bibinfo
  {journal} {Nat. Physics}\ }\textbf {\bibinfo {volume} {16}},\ \bibinfo
  {pages} {247} (\bibinfo {year} {2020})}\BibitemShut {NoStop}%
\bibitem [{\citenamefont {Haroche}\ \emph {et~al.}(2020)\citenamefont
  {Haroche}, \citenamefont {Brune},\ and\ \citenamefont {Raimond}}]{SQC_3}%
  \BibitemOpen
  \bibfield  {author} {\bibinfo {author} {\bibfnamefont {S.}~\bibnamefont
  {Haroche}}, \bibinfo {author} {\bibfnamefont {M.}~\bibnamefont {Brune}},\
  and\ \bibinfo {author} {\bibfnamefont {J.~M.}\ \bibnamefont {Raimond}},\
  }\bibfield  {title} {\bibinfo {title} {From cavity to circuit quantum
  electrodynamics},\ }\href {https://doi.org/10.1038/s41567-020-0812-1}
  {\bibfield  {journal} {\bibinfo  {journal} {Nat. Physics}\ }\textbf {\bibinfo
  {volume} {16}},\ \bibinfo {pages} {243} (\bibinfo {year} {2020})}\BibitemShut
  {NoStop}%
\bibitem [{\citenamefont {Blais}\ \emph {et~al.}(2004)\citenamefont {Blais},
  \citenamefont {Huang}, \citenamefont {Wallraff}, \citenamefont {Girvin},\
  and\ \citenamefont {Schoelkopf}}]{SQC_4}%
  \BibitemOpen
  \bibfield  {author} {\bibinfo {author} {\bibfnamefont {A.}~\bibnamefont
  {Blais}}, \bibinfo {author} {\bibfnamefont {R.-S.}\ \bibnamefont {Huang}},
  \bibinfo {author} {\bibfnamefont {A.}~\bibnamefont {Wallraff}}, \bibinfo
  {author} {\bibfnamefont {S.~M.}\ \bibnamefont {Girvin}},\ and\ \bibinfo
  {author} {\bibfnamefont {R.~J.}\ \bibnamefont {Schoelkopf}},\ }\bibfield
  {title} {\bibinfo {title} {Cavity quantum electrodynamics for superconducting
  electrical circuits: An architecture for quantum computation},\ }\href
  {https://doi.org/10.1103/PhysRevA.69.062320} {\bibfield  {journal} {\bibinfo
  {journal} {Phys. Rev. A}\ }\textbf {\bibinfo {volume} {69}},\ \bibinfo
  {pages} {062320} (\bibinfo {year} {2004})}\BibitemShut {NoStop}%
\bibitem [{\citenamefont {Scigliuzzo}\ \emph {et~al.}(2022)\citenamefont
  {Scigliuzzo}, \citenamefont {Calaj\`o}, \citenamefont {Ciccarello},
  \citenamefont {Perez~Lozano}, \citenamefont {Bengtsson}, \citenamefont
  {Scarlino}, \citenamefont {Wallraff}, \citenamefont {Chang}, \citenamefont
  {Delsing},\ and\ \citenamefont {Gasparinetti}}]{SQC_CRW_1}%
  \BibitemOpen
  \bibfield  {author} {\bibinfo {author} {\bibfnamefont {M.}~\bibnamefont
  {Scigliuzzo}}, \bibinfo {author} {\bibfnamefont {G.}~\bibnamefont
  {Calaj\`o}}, \bibinfo {author} {\bibfnamefont {F.}~\bibnamefont
  {Ciccarello}}, \bibinfo {author} {\bibfnamefont {D.}~\bibnamefont
  {Perez~Lozano}}, \bibinfo {author} {\bibfnamefont {A.}~\bibnamefont
  {Bengtsson}}, \bibinfo {author} {\bibfnamefont {P.}~\bibnamefont {Scarlino}},
  \bibinfo {author} {\bibfnamefont {A.}~\bibnamefont {Wallraff}}, \bibinfo
  {author} {\bibfnamefont {D.}~\bibnamefont {Chang}}, \bibinfo {author}
  {\bibfnamefont {P.}~\bibnamefont {Delsing}},\ and\ \bibinfo {author}
  {\bibfnamefont {S.}~\bibnamefont {Gasparinetti}},\ }\bibfield  {title}
  {\bibinfo {title} {Controlling atom-photon bound states in an array of
  josephson-junction resonators},\ }\href
  {https://doi.org/10.1103/PhysRevX.12.031036} {\bibfield  {journal} {\bibinfo
  {journal} {Phys. Rev. X}\ }\textbf {\bibinfo {volume} {12}},\ \bibinfo
  {pages} {031036} (\bibinfo {year} {2022})}\BibitemShut {NoStop}%
\bibitem [{\citenamefont {Wang}\ \emph
  {et~al.}(2021{\natexlab{b}})\citenamefont {Wang}, \citenamefont {Liu},
  \citenamefont {Kockum}, \citenamefont {Li},\ and\ \citenamefont
  {Nori}}]{SQC_CRW_2}%
  \BibitemOpen
  \bibfield  {author} {\bibinfo {author} {\bibfnamefont {X.}~\bibnamefont
  {Wang}}, \bibinfo {author} {\bibfnamefont {T.}~\bibnamefont {Liu}}, \bibinfo
  {author} {\bibfnamefont {A.~F.}\ \bibnamefont {Kockum}}, \bibinfo {author}
  {\bibfnamefont {H.-R.}\ \bibnamefont {Li}},\ and\ \bibinfo {author}
  {\bibfnamefont {F.}~\bibnamefont {Nori}},\ }\bibfield  {title} {\bibinfo
  {title} {Tunable chiral bound states with giant atoms},\ }\href
  {https://doi.org/10.1103/PhysRevLett.126.043602} {\bibfield  {journal}
  {\bibinfo  {journal} {Phys. Rev. Lett.}\ }\textbf {\bibinfo {volume} {126}},\
  \bibinfo {pages} {043602} (\bibinfo {year} {2021}{\natexlab{b}})}\BibitemShut
  {NoStop}%
\bibitem [{\citenamefont {Yu}\ \emph {et~al.}(2021)\citenamefont {Yu},
  \citenamefont {Wang},\ and\ \citenamefont {Wu}}]{SQC_CRW_3}%
  \BibitemOpen
  \bibfield  {author} {\bibinfo {author} {\bibfnamefont {H.}~\bibnamefont
  {Yu}}, \bibinfo {author} {\bibfnamefont {Z.}~\bibnamefont {Wang}},\ and\
  \bibinfo {author} {\bibfnamefont {J.-H.}\ \bibnamefont {Wu}},\ }\bibfield
  {title} {\bibinfo {title} {Entanglement preparation and nonreciprocal
  excitation evolution in giant atoms by controllable dissipation and
  coupling},\ }\href {https://doi.org/10.1103/PhysRevA.104.013720} {\bibfield
  {journal} {\bibinfo  {journal} {Phys. Rev. A}\ }\textbf {\bibinfo {volume}
  {104}},\ \bibinfo {pages} {013720} (\bibinfo {year} {2021})}\BibitemShut
  {NoStop}%
\bibitem [{\citenamefont {You}\ \emph {et~al.}(2007)\citenamefont {You},
  \citenamefont {Liu}, \citenamefont {Sun},\ and\ \citenamefont
  {Nori}}]{SQC_GA_1}%
  \BibitemOpen
  \bibfield  {author} {\bibinfo {author} {\bibfnamefont {J.~Q.}\ \bibnamefont
  {You}}, \bibinfo {author} {\bibfnamefont {Y.-x.}\ \bibnamefont {Liu}},
  \bibinfo {author} {\bibfnamefont {C.~P.}\ \bibnamefont {Sun}},\ and\ \bibinfo
  {author} {\bibfnamefont {F.}~\bibnamefont {Nori}},\ }\bibfield  {title}
  {\bibinfo {title} {Persistent single-photon production by tunable on-chip
  micromaser with a superconducting quantum circuit},\ }\href
  {https://doi.org/10.1103/PhysRevB.75.104516} {\bibfield  {journal} {\bibinfo
  {journal} {Phys. Rev. B}\ }\textbf {\bibinfo {volume} {75}},\ \bibinfo
  {pages} {104516} (\bibinfo {year} {2007})}\BibitemShut {NoStop}%
\bibitem [{\citenamefont {Wang}\ \emph {et~al.}(2022)\citenamefont {Wang},
  \citenamefont {Yang}, \citenamefont {Chen}, \citenamefont {Li}, \citenamefont
  {Shui}, \citenamefont {Li},\ and\ \citenamefont {Wu}}]{SQC_GA_2}%
  \BibitemOpen
  \bibfield  {author} {\bibinfo {author} {\bibfnamefont {X.}~\bibnamefont
  {Wang}}, \bibinfo {author} {\bibfnamefont {W.-X.}\ \bibnamefont {Yang}},
  \bibinfo {author} {\bibfnamefont {A.-X.}\ \bibnamefont {Chen}}, \bibinfo
  {author} {\bibfnamefont {L.}~\bibnamefont {Li}}, \bibinfo {author}
  {\bibfnamefont {T.}~\bibnamefont {Shui}}, \bibinfo {author} {\bibfnamefont
  {X.}~\bibnamefont {Li}},\ and\ \bibinfo {author} {\bibfnamefont
  {Z.}~\bibnamefont {Wu}},\ }\bibfield  {title} {\bibinfo {title}
  {Phase-modulated single-photon nonreciprocal transport and directional router
  in a waveguide–cavity–emitter system beyond the chiral coupling},\ }\href
  {https://doi.org/10.1088/2058-9565/ac4425} {\bibfield  {journal} {\bibinfo
  {journal} {Quantum Sci. Technol.}\ }\textbf {\bibinfo {volume} {7}},\
  \bibinfo {pages} {015025} (\bibinfo {year} {2022})}\BibitemShut {NoStop}%
\bibitem [{\citenamefont {Du}\ \emph {et~al.}(2022)\citenamefont {Du},
  \citenamefont {Zhang}, \citenamefont {Wu}, \citenamefont {Kockum},\ and\
  \citenamefont {Li}}]{SQC_GA_3}%
  \BibitemOpen
  \bibfield  {author} {\bibinfo {author} {\bibfnamefont {L.}~\bibnamefont
  {Du}}, \bibinfo {author} {\bibfnamefont {Y.}~\bibnamefont {Zhang}}, \bibinfo
  {author} {\bibfnamefont {J.-H.}\ \bibnamefont {Wu}}, \bibinfo {author}
  {\bibfnamefont {A.~F.}\ \bibnamefont {Kockum}},\ and\ \bibinfo {author}
  {\bibfnamefont {Y.}~\bibnamefont {Li}},\ }\bibfield  {title} {\bibinfo
  {title} {Giant atoms in a synthetic frequency dimension},\ }\href
  {https://doi.org/10.1103/PhysRevLett.128.223602} {\bibfield  {journal}
  {\bibinfo  {journal} {Phys. Rev. Lett.}\ }\textbf {\bibinfo {volume} {128}},\
  \bibinfo {pages} {223602} (\bibinfo {year} {2022})}\BibitemShut {NoStop}%
\bibitem [{\citenamefont {Frisk~Kockum}(2021)}]{giant_atom_1}%
  \BibitemOpen
  \bibfield  {author} {\bibinfo {author} {\bibfnamefont {A.}~\bibnamefont
  {Frisk~Kockum}},\ }\bibfield  {title} {\bibinfo {title} {Quantum optics with
  giant atoms---the first five years},\ }in\ \href
  {https://doi.org/https://doi.org/10.1007/978-981-15-5191-8_12} {\emph
  {\bibinfo {booktitle} {International Symposium on Mathematics, Quantum
  Theory, and Cryptography}}},\ \bibinfo {editor} {edited by\ \bibinfo {editor}
  {\bibfnamefont {T.}~\bibnamefont {Takagi}}, \bibinfo {editor} {\bibfnamefont
  {M.}~\bibnamefont {Wakayama}}, \bibinfo {editor} {\bibfnamefont
  {K.}~\bibnamefont {Tanaka}}, \bibinfo {editor} {\bibfnamefont
  {N.}~\bibnamefont {Kunihiro}}, \bibinfo {editor} {\bibfnamefont
  {K.}~\bibnamefont {Kimoto}},\ and\ \bibinfo {editor} {\bibfnamefont
  {Y.}~\bibnamefont {Ikematsu}}}\ (\bibinfo  {publisher} {Springer Singapore},\
  \bibinfo {address} {Singapore},\ \bibinfo {year} {2021})\ pp.\ \bibinfo
  {pages} {125--146}\BibitemShut {NoStop}%
\bibitem [{\citenamefont {Kannan}\ \emph {et~al.}(2020)\citenamefont {Kannan},
  \citenamefont {Ruckriegel}, \citenamefont {Campbell}, \citenamefont
  {Frisk~Kockum}, \citenamefont {Braumüller}, \citenamefont {Kim},
  \citenamefont {Kjaergaard}, \citenamefont {Krantz}, \citenamefont {Melville},
  \citenamefont {Niedzielski}, \citenamefont {Vepsäläinen}, \citenamefont
  {Winik}, \citenamefont {Yoder}, \citenamefont {Nori}, \citenamefont
  {Orlando}, \citenamefont {Gustavsson},\ and\ \citenamefont
  {Oliver}}]{giant_atom_2}%
  \BibitemOpen
  \bibfield  {author} {\bibinfo {author} {\bibfnamefont {B.}~\bibnamefont
  {Kannan}}, \bibinfo {author} {\bibfnamefont {M.}~\bibnamefont {Ruckriegel}},
  \bibinfo {author} {\bibfnamefont {D.}~\bibnamefont {Campbell}}, \bibinfo
  {author} {\bibfnamefont {A.}~\bibnamefont {Frisk~Kockum}}, \bibinfo {author}
  {\bibfnamefont {J.}~\bibnamefont {Braumüller}}, \bibinfo {author}
  {\bibfnamefont {D.}~\bibnamefont {Kim}}, \bibinfo {author} {\bibfnamefont
  {M.}~\bibnamefont {Kjaergaard}}, \bibinfo {author} {\bibfnamefont
  {P.}~\bibnamefont {Krantz}}, \bibinfo {author} {\bibfnamefont
  {A.}~\bibnamefont {Melville}}, \bibinfo {author} {\bibfnamefont
  {B.}~\bibnamefont {Niedzielski}}, \bibinfo {author} {\bibfnamefont
  {A.}~\bibnamefont {Vepsäläinen}}, \bibinfo {author} {\bibfnamefont
  {R.}~\bibnamefont {Winik}}, \bibinfo {author} {\bibfnamefont
  {J.}~\bibnamefont {Yoder}}, \bibinfo {author} {\bibfnamefont
  {F.}~\bibnamefont {Nori}}, \bibinfo {author} {\bibfnamefont {T.}~\bibnamefont
  {Orlando}}, \bibinfo {author} {\bibfnamefont {S.}~\bibnamefont
  {Gustavsson}},\ and\ \bibinfo {author} {\bibfnamefont {W.}~\bibnamefont
  {Oliver}},\ }\bibfield  {title} {\bibinfo {title} {Waveguide quantum
  electrodynamics with superconducting artificial giant atoms},\ }\href
  {https://doi.org/10.1038/s41586-020-2529-9} {\bibfield  {journal} {\bibinfo
  {journal} {Nature}\ }\textbf {\bibinfo {volume} {583}},\ \bibinfo {pages}
  {775} (\bibinfo {year} {2020})}\BibitemShut {NoStop}%
\bibitem [{\citenamefont {Vadiraj}\ \emph {et~al.}(2021)\citenamefont
  {Vadiraj}, \citenamefont {Ask}, \citenamefont {McConkey}, \citenamefont
  {Nsanzineza}, \citenamefont {Chang}, \citenamefont {Kockum},\ and\
  \citenamefont {Wilson}}]{giant_atom_engineering}%
  \BibitemOpen
  \bibfield  {author} {\bibinfo {author} {\bibfnamefont {A.~M.}\ \bibnamefont
  {Vadiraj}}, \bibinfo {author} {\bibfnamefont {A.}~\bibnamefont {Ask}},
  \bibinfo {author} {\bibfnamefont {T.~G.}\ \bibnamefont {McConkey}}, \bibinfo
  {author} {\bibfnamefont {I.}~\bibnamefont {Nsanzineza}}, \bibinfo {author}
  {\bibfnamefont {C.~W.~S.}\ \bibnamefont {Chang}}, \bibinfo {author}
  {\bibfnamefont {A.~F.}\ \bibnamefont {Kockum}},\ and\ \bibinfo {author}
  {\bibfnamefont {C.~M.}\ \bibnamefont {Wilson}},\ }\bibfield  {title}
  {\bibinfo {title} {Engineering the level structure of a giant artificial atom
  in waveguide quantum electrodynamics},\ }\href
  {https://doi.org/10.1103/PhysRevA.103.023710} {\bibfield  {journal} {\bibinfo
   {journal} {Phys. Rev. A}\ }\textbf {\bibinfo {volume} {103}},\ \bibinfo
  {pages} {023710} (\bibinfo {year} {2021})}\BibitemShut {NoStop}%
\bibitem [{\citenamefont {Andersson}\ \emph {et~al.}(2020)\citenamefont
  {Andersson}, \citenamefont {Ekstr\"om},\ and\ \citenamefont
  {Delsing}}]{giant_atom_3}%
  \BibitemOpen
  \bibfield  {author} {\bibinfo {author} {\bibfnamefont {G.}~\bibnamefont
  {Andersson}}, \bibinfo {author} {\bibfnamefont {M.~K.}\ \bibnamefont
  {Ekstr\"om}},\ and\ \bibinfo {author} {\bibfnamefont {P.}~\bibnamefont
  {Delsing}},\ }\bibfield  {title} {\bibinfo {title} {Electromagnetically
  induced acoustic transparency with a superconducting circuit},\ }\href
  {https://doi.org/10.1103/PhysRevLett.124.240402} {\bibfield  {journal}
  {\bibinfo  {journal} {Phys. Rev. Lett.}\ }\textbf {\bibinfo {volume} {124}},\
  \bibinfo {pages} {240402} (\bibinfo {year} {2020})}\BibitemShut {NoStop}%
\bibitem [{\citenamefont {Zhang}\ \emph {et~al.}(2023)\citenamefont {Zhang},
  \citenamefont {Liu}, \citenamefont {Gong},\ and\ \citenamefont
  {Wang}}]{giant_atom_crw_sistem}%
  \BibitemOpen
  \bibfield  {author} {\bibinfo {author} {\bibfnamefont {X.}~\bibnamefont
  {Zhang}}, \bibinfo {author} {\bibfnamefont {C.}~\bibnamefont {Liu}}, \bibinfo
  {author} {\bibfnamefont {Z.}~\bibnamefont {Gong}},\ and\ \bibinfo {author}
  {\bibfnamefont {Z.}~\bibnamefont {Wang}},\ }\bibfield  {title} {\bibinfo
  {title} {Quantum interference and controllable magic cavity qed via a giant
  atom in a coupled resonator waveguide},\ }\href
  {https://doi.org/10.1103/PhysRevA.108.013704} {\bibfield  {journal} {\bibinfo
   {journal} {Phys. Rev. A}\ }\textbf {\bibinfo {volume} {108}},\ \bibinfo
  {pages} {013704} (\bibinfo {year} {2023})}\BibitemShut {NoStop}%
\bibitem [{\citenamefont {Chen}\ \emph {et~al.}(2022)\citenamefont {Chen},
  \citenamefont {Du}, \citenamefont {Guo}, \citenamefont {Wang}, \citenamefont
  {Zhang}, \citenamefont {Li},\ and\ \citenamefont {Wu}}]{giant_atom_router_1}%
  \BibitemOpen
  \bibfield  {author} {\bibinfo {author} {\bibfnamefont {Y.-T.}\ \bibnamefont
  {Chen}}, \bibinfo {author} {\bibfnamefont {L.}~\bibnamefont {Du}}, \bibinfo
  {author} {\bibfnamefont {L.}~\bibnamefont {Guo}}, \bibinfo {author}
  {\bibfnamefont {Z.}~\bibnamefont {Wang}}, \bibinfo {author} {\bibfnamefont
  {Y.}~\bibnamefont {Zhang}}, \bibinfo {author} {\bibfnamefont
  {Y.}~\bibnamefont {Li}},\ and\ \bibinfo {author} {\bibfnamefont {J.-H.}\
  \bibnamefont {Wu}},\ }\bibfield  {title} {\bibinfo {title} {Nonreciprocal and
  chiral single-photon scattering for giant atoms},\ }\href
  {https://doi.org/10.1038/s42005-022-00991-3} {\bibfield  {journal} {\bibinfo
  {journal} {Commun. Phys.}\ }\textbf {\bibinfo {volume} {5}},\ \bibinfo
  {pages} {215} (\bibinfo {year} {2022})}\BibitemShut {NoStop}%
\bibitem [{\citenamefont {Du}\ \emph {et~al.}(2021)\citenamefont {Du},
  \citenamefont {Chen},\ and\ \citenamefont {Li}}]{giant_atom_router_2}%
  \BibitemOpen
  \bibfield  {author} {\bibinfo {author} {\bibfnamefont {L.}~\bibnamefont
  {Du}}, \bibinfo {author} {\bibfnamefont {Y.-T.}\ \bibnamefont {Chen}},\ and\
  \bibinfo {author} {\bibfnamefont {Y.}~\bibnamefont {Li}},\ }\bibfield
  {title} {\bibinfo {title} {Nonreciprocal frequency conversion with chiral
  $\mathrm{\ensuremath{\Lambda}}$-type atoms},\ }\href
  {https://doi.org/10.1103/PhysRevResearch.3.043226} {\bibfield  {journal}
  {\bibinfo  {journal} {Phys. Rev. Res.}\ }\textbf {\bibinfo {volume} {3}},\
  \bibinfo {pages} {043226} (\bibinfo {year} {2021})}\BibitemShut {NoStop}%
\bibitem [{\citenamefont {Zhao}\ and\ \citenamefont
  {Wang}(2020)}]{giant_atom_router_3}%
  \BibitemOpen
  \bibfield  {author} {\bibinfo {author} {\bibfnamefont {W.}~\bibnamefont
  {Zhao}}\ and\ \bibinfo {author} {\bibfnamefont {Z.}~\bibnamefont {Wang}},\
  }\bibfield  {title} {\bibinfo {title} {Single-photon scattering and bound
  states in an atom-waveguide system with two or multiple coupling points},\
  }\href {https://doi.org/10.1103/PhysRevA.101.053855} {\bibfield  {journal}
  {\bibinfo  {journal} {Phys. Rev. A}\ }\textbf {\bibinfo {volume} {101}},\
  \bibinfo {pages} {053855} (\bibinfo {year} {2020})}\BibitemShut {NoStop}%
\bibitem [{\citenamefont {Morichetti}\ \emph {et~al.}(2012)\citenamefont
  {Morichetti}, \citenamefont {Ferrari}, \citenamefont {Canciamilla},\ and\
  \citenamefont {Melloni}}]{crw_hamiltonian_1}%
  \BibitemOpen
  \bibfield  {author} {\bibinfo {author} {\bibfnamefont {F.}~\bibnamefont
  {Morichetti}}, \bibinfo {author} {\bibfnamefont {C.}~\bibnamefont {Ferrari}},
  \bibinfo {author} {\bibfnamefont {A.}~\bibnamefont {Canciamilla}},\ and\
  \bibinfo {author} {\bibfnamefont {A.}~\bibnamefont {Melloni}},\ }\bibfield
  {title} {\bibinfo {title} {The first decade of coupled resonator optical
  waveguides: bringing slow light to applications},\ }\href
  {https://doi.org/https://doi.org/10.1002/lpor.201100018} {\bibfield
  {journal} {\bibinfo  {journal} {Laser Photonics Rev.}\ }\textbf {\bibinfo
  {volume} {6}},\ \bibinfo {pages} {74} (\bibinfo {year} {2012})}\BibitemShut
  {NoStop}%
\bibitem [{\citenamefont {Yariv}\ \emph {et~al.}(1999)\citenamefont {Yariv},
  \citenamefont {Xu}, \citenamefont {Lee},\ and\ \citenamefont
  {Scherer}}]{crw_hamiltonian_2}%
  \BibitemOpen
  \bibfield  {author} {\bibinfo {author} {\bibfnamefont {A.}~\bibnamefont
  {Yariv}}, \bibinfo {author} {\bibfnamefont {Y.}~\bibnamefont {Xu}}, \bibinfo
  {author} {\bibfnamefont {R.~K.}\ \bibnamefont {Lee}},\ and\ \bibinfo {author}
  {\bibfnamefont {A.}~\bibnamefont {Scherer}},\ }\bibfield  {title} {\bibinfo
  {title} {Coupled-resonator optical waveguide:?a proposal and analysis},\
  }\href {https://doi.org/10.1364/OL.24.000711} {\bibfield  {journal} {\bibinfo
   {journal} {Opt. Lett.}\ }\textbf {\bibinfo {volume} {24}},\ \bibinfo {pages}
  {711} (\bibinfo {year} {1999})}\BibitemShut {NoStop}%
\bibitem [{\citenamefont {Hartmann}\ \emph {et~al.}(2008)\citenamefont
  {Hartmann}, \citenamefont {Brandao},\ and\ \citenamefont
  {Plenio}}]{crw_hamiltonian_3}%
  \BibitemOpen
  \bibfield  {author} {\bibinfo {author} {\bibfnamefont {M.~J.}\ \bibnamefont
  {Hartmann}}, \bibinfo {author} {\bibfnamefont {F.~G.}\ \bibnamefont
  {Brandao}},\ and\ \bibinfo {author} {\bibfnamefont {M.~B.}\ \bibnamefont
  {Plenio}},\ }\bibfield  {title} {\bibinfo {title} {Quantum many-body
  phenomena in coupled cavity arrays},\ }\href
  {https://doi.org/https://doi.org/10.1002/lpor.200810046} {\bibfield
  {journal} {\bibinfo  {journal} {Laser Photonics Rev.}\ }\textbf {\bibinfo
  {volume} {2}},\ \bibinfo {pages} {527} (\bibinfo {year} {2008})}\BibitemShut
  {NoStop}%
\bibitem [{\citenamefont {Liu}\ \emph {et~al.}(2005)\citenamefont {Liu},
  \citenamefont {You}, \citenamefont {Wei}, \citenamefont {Sun},\ and\
  \citenamefont {Nori}}]{Selection_Rule}%
  \BibitemOpen
  \bibfield  {author} {\bibinfo {author} {\bibfnamefont {Y.-x.}\ \bibnamefont
  {Liu}}, \bibinfo {author} {\bibfnamefont {J.~Q.}\ \bibnamefont {You}},
  \bibinfo {author} {\bibfnamefont {L.~F.}\ \bibnamefont {Wei}}, \bibinfo
  {author} {\bibfnamefont {C.~P.}\ \bibnamefont {Sun}},\ and\ \bibinfo {author}
  {\bibfnamefont {F.}~\bibnamefont {Nori}},\ }\bibfield  {title} {\bibinfo
  {title} {Optical selection rules and phase-dependent adiabatic state control
  in a superconducting quantum circuit},\ }\href
  {https://doi.org/10.1103/PhysRevLett.95.087001} {\bibfield  {journal}
  {\bibinfo  {journal} {Phys. Rev. Lett.}\ }\textbf {\bibinfo {volume} {95}},\
  \bibinfo {pages} {087001} (\bibinfo {year} {2005})}\BibitemShut {NoStop}%
\bibitem [{\citenamefont {Yan}\ \emph {et~al.}(2023)\citenamefont {Yan},
  \citenamefont {Li}, \citenamefont {Xu}, \citenamefont {Zhang}, \citenamefont
  {Ma},\ and\ \citenamefont {Zou}}]{destructive_interference_1}%
  \BibitemOpen
  \bibfield  {author} {\bibinfo {author} {\bibfnamefont {C.-H.}\ \bibnamefont
  {Yan}}, \bibinfo {author} {\bibfnamefont {M.}~\bibnamefont {Li}}, \bibinfo
  {author} {\bibfnamefont {X.-B.}\ \bibnamefont {Xu}}, \bibinfo {author}
  {\bibfnamefont {Y.-L.}\ \bibnamefont {Zhang}}, \bibinfo {author}
  {\bibfnamefont {X.-Y.}\ \bibnamefont {Ma}},\ and\ \bibinfo {author}
  {\bibfnamefont {C.-L.}\ \bibnamefont {Zou}},\ }\bibfield  {title} {\bibinfo
  {title} {Unidirectional propagation of single photons realized by a scatterer
  coupled to whispering-gallery-mode microresonators},\ }\href
  {https://doi.org/10.1103/PhysRevA.107.033713} {\bibfield  {journal} {\bibinfo
   {journal} {Phys. Rev. A}\ }\textbf {\bibinfo {volume} {107}},\ \bibinfo
  {pages} {033713} (\bibinfo {year} {2023})}\BibitemShut {NoStop}%
\bibitem [{\citenamefont {Svela}\ \emph {et~al.}(2020)\citenamefont {Svela},
  \citenamefont {Silver}, \citenamefont {Del~Bino}, \citenamefont {Zhang},
  \citenamefont {Woodley}, \citenamefont {Vanner},\ and\ \citenamefont
  {Del'Haye}}]{destructive_interference_2}%
  \BibitemOpen
  \bibfield  {author} {\bibinfo {author} {\bibfnamefont {A.~{\O}.}\
  \bibnamefont {Svela}}, \bibinfo {author} {\bibfnamefont {J.~M.}\ \bibnamefont
  {Silver}}, \bibinfo {author} {\bibfnamefont {L.}~\bibnamefont {Del~Bino}},
  \bibinfo {author} {\bibfnamefont {S.}~\bibnamefont {Zhang}}, \bibinfo
  {author} {\bibfnamefont {M.~T.~M.}\ \bibnamefont {Woodley}}, \bibinfo
  {author} {\bibfnamefont {M.~R.}\ \bibnamefont {Vanner}},\ and\ \bibinfo
  {author} {\bibfnamefont {P.}~\bibnamefont {Del'Haye}},\ }\bibfield  {title}
  {\bibinfo {title} {Coherent suppression of backscattering in optical
  microresonators},\ }\href {https://doi.org/10.1038/s41377-020-00440-2}
  {\bibfield  {journal} {\bibinfo  {journal} {Light Sci. Appl.}\ }\textbf
  {\bibinfo {volume} {9}},\ \bibinfo {pages} {204} (\bibinfo {year}
  {2020})}\BibitemShut {NoStop}%
\bibitem [{\citenamefont {Wang}\ \emph {et~al.}(2020)\citenamefont {Wang},
  \citenamefont {Jiang}, \citenamefont {Zhao}, \citenamefont {Zhang},
  \citenamefont {Hsu}, \citenamefont {Peng}, \citenamefont {Stone},
  \citenamefont {Jiang},\ and\ \citenamefont
  {Yang}}]{destructive_interference_3}%
  \BibitemOpen
  \bibfield  {author} {\bibinfo {author} {\bibfnamefont {C.}~\bibnamefont
  {Wang}}, \bibinfo {author} {\bibfnamefont {X.}~\bibnamefont {Jiang}},
  \bibinfo {author} {\bibfnamefont {G.}~\bibnamefont {Zhao}}, \bibinfo {author}
  {\bibfnamefont {M.}~\bibnamefont {Zhang}}, \bibinfo {author} {\bibfnamefont
  {C.~W.}\ \bibnamefont {Hsu}}, \bibinfo {author} {\bibfnamefont
  {B.}~\bibnamefont {Peng}}, \bibinfo {author} {\bibfnamefont {A.~D.}\
  \bibnamefont {Stone}}, \bibinfo {author} {\bibfnamefont {L.}~\bibnamefont
  {Jiang}},\ and\ \bibinfo {author} {\bibfnamefont {L.}~\bibnamefont {Yang}},\
  }\bibfield  {title} {\bibinfo {title} {Electromagnetically induced
  transparency at a chiral exceptional point},\ }\href
  {https://doi.org/10.1038/s41567-019-0746-7} {\bibfield  {journal} {\bibinfo
  {journal} {Nat. Phys.}\ }\textbf {\bibinfo {volume} {16}},\ \bibinfo {pages}
  {334} (\bibinfo {year} {2020})}\BibitemShut {NoStop}%
\bibitem [{\citenamefont {Wang}\ \emph {et~al.}(2016)\citenamefont {Wang},
  \citenamefont {Zhang}, \citenamefont {Zhang}, \citenamefont {Zou},\ and\
  \citenamefont {Schwingenschlogl}}]{destructive_interference_4}%
  \BibitemOpen
  \bibfield  {author} {\bibinfo {author} {\bibfnamefont {Y.}~\bibnamefont
  {Wang}}, \bibinfo {author} {\bibfnamefont {Y.}~\bibnamefont {Zhang}},
  \bibinfo {author} {\bibfnamefont {Q.}~\bibnamefont {Zhang}}, \bibinfo
  {author} {\bibfnamefont {B.}~\bibnamefont {Zou}},\ and\ \bibinfo {author}
  {\bibfnamefont {U.}~\bibnamefont {Schwingenschlogl}},\ }\bibfield  {title}
  {\bibinfo {title} {Dynamics of single photon transport in a one-dimensional
  waveguide two-point coupled with a jaynes-cummings system},\ }\href
  {https://doi.org/10.1038/srep33867} {\bibfield  {journal} {\bibinfo
  {journal} {Sci. Rep.}\ }\textbf {\bibinfo {volume} {6}},\ \bibinfo {pages}
  {33867} (\bibinfo {year} {2016})}\BibitemShut {NoStop}%
\end{thebibliography}%

\end{document}